\documentclass[12pt,a4paper]{article}
\usepackage{amsfonts}
\newcommand{\Spc}{\mbox{Spin}^c(M)\times_{\mbox{\scriptsize Spin}^c(n)}}
\newcommand{\Spt}{\mbox{Spin}(M)\times_{\mbox{\scriptsize Spin}(n)}}
\newcommand{\mfS}{\mathfrak{S}}
\newcommand{\mfs}{\mathfrak{s}}

\newcommand{\Ga}{{\mit\Gamma}}
\newcommand{\DIV}{\mbox{div}\,}
\newcommand{\dg}{|\mbox{det}\,g|^{1/2}}
\newcommand{\dgm}{|\mbox{det}\,g|^{-1/2}}

\newcommand{\Tr}{\mbox{Tr}\,}
\newcommand{\pa}[1]{(-1)^{|#1|}}
\newcommand{\Ker}{\mbox{ker}\,}
\newcommand{\Ima}{\mbox{im}\,}
\newcommand{\RR}{\mathbb{R}}
\newcommand{\CC}{\mathbb{C}}
\newcommand{\ZZ}{\mathbb{Z}}

\newcommand{\D}{I\kern-3.5pt D}
\newcommand{\F}{{I\kern-3.5pt F}}
\newcommand{\R}{\mathfrak{R}}
\newcommand{\io}{\iota}

\newcommand{\ep}{\epsilon}
\newcommand{\frc}[2]{{\textstyle\frac{#1}{#2}}}
\newcommand{\one}{1\kern-3pt{\rm l}}
\newcommand{\hf}{\frc{1}{2}}
\newcommand{\bw}{{\textstyle\bigwedge}}

\newcommand{\Om}{\Omega}
\newcommand{\om}{\omega}
\newcommand{\br}[1]{\langle #1\rangle}
\newcommand{\arr}[1]{\,\,\smash{\mathop{\longrightarrow}\limits^{#1}}\,\,}
\newcommand{\llra}{\relbar\joinrel\relbar\joinrel\longrightarrow}
\newcommand{\mapr}[1]{\smash{\mathop{\llra}\limits^{#1}}}
\newcommand{\mapd}[1]{\Big\downarrow\rlap{$\vcenter{\hbox{$\scriptstyle#1$}}$}}
\newcommand{\sco}[1]{\,[\![\,{#1}\,]\!]\,}
\newcommand{\nein}{\in\kern-0.8em/\,}
\newcommand{\G}[1]{\Gamma(#1)}
\newcommand{\E}{\mathfrak{E}}
\newcommand{\V}{\mathfrak{V}}
\newcommand{\Ff}{\mathfrak{F}}
\newcommand{\pd}{\partial}
\newcommand{\End}{\mbox{End}\,}
\newcommand{\gr}{\mbox{grad}\,}
\newcommand{\nsl}{\nabla\kern-9pt/\kern3pt}
\newcommand{\psl}{\partial\kern-6pt/}
\newcommand{\asl}{A\kern-6pt/}
\newcommand{\Dsl}{\D\kern-8pt/\kern2pt}
\newcommand{\Fsl}{F\kern-7pt/\kern1pt}
\newcommand{\Rsl}{R\kern-7pt/\kern1pt}
\newcommand{\skt}{{\hat\otimes}}

\begin{document}
\vspace*{-2cm}
\rightline{October 1999}
\vspace{2cm}
\begin{center}
 {\LARGE\bf Generalized Dirac Operators\\[3mm]
  and Superconnections}\\[15mm]
  {\large G.\ Roepstorff and Ch.\ Vehns}\\
  Institute for Theoretical Physics\\
  RWTH Aachen\\
  D-52062 Aachen, Germany\\
  e-mail:\\
  roep@physik.rwth-aachen.de\\
  vehns@physik.rwth-aachen.de\\[2cm]
  {\bf Abstract}
\end{center}
\begin{quote}
Motivated by the supersymmetric version of Dirac's theory, chiral models
in field theory, and the quest of a geometric fundament for the Standard 
Model, we describe an approach to differential geometry of vector bundles on
(semi-)Riemannian manifolds based on the concepts of 
superspaces, superalgebras, superconnections, and generalized Dirac operators. 
We stay within the realm of {\em commutative geometry}.
\end{quote}

\section{Introduction}

This is the second part of a series of articles devoted to the theory of
fermions in a setting which emphasizes the $\ZZ_2$-graded structure of
generalized (multi-component) Dirac fields or, mathematically speaking, of
Clifford modules and, even more importantly, of operators acting on them. 
While in [1] we studied the purely algebraic aspects, we shall
now focus on questions pertaining to the differential geometry of 
superbundles which, at the same time, are Clifford modules. In doing so,
we try to keep close to the formulation and notational convention of
the textbook by Berline, Getzler, and Vergne [2].

Crudely speaking, superbundles have superspaces as their fibers.
By ``superspace'' we mean any vector space $V$ which is $\ZZ_2$-graded:
$$
             V=V^+\oplus V^-,\qquad \mbox{dim}\,V^+=\mbox{dim}\,V^-\,.
$$
Normally, one interpretes $\ZZ_2$ as the additive group $\ZZ/2\ZZ$ and
calls $|v|=0,1$ the {\em degree\/} of the vector $v\in V$, while 
in physics we prefer to work with parities:
$$
                 \mbox{par}(v)=(-1)^{|v|}\,.
$$
Endomorphisms of a superspace naturally form a superalgebra. Operators are
said to possess negative (positive) parity if they are (not) parity changing.
For brevity, we call them odd (even) operators.
When dealing with superalgebras, care has to be exercised with regard to the
tensor product. We exclusively use the `skew tensor product' indicated by
the symbol $\skt$. See [1] for details. Basic relations are formulated in
terms of supercommutators for which we use special brackets:
$$
            \sco{a,b}=\cases{ab+ba& if $a$ and $b$ are odd\cr
                             ab-ba& otherwise.\cr}
$$
The new ingredient, fundamental to both Riemannian geometry and field theory,
is the concept of (generalized) Dirac operators and their associated
Laplacians. The guiding principle is this: Dirac operators are quantized
superconnections and Laplacians arise as squares of Dirac operators.
Here, ``quantization'' enters as a mathematical concept referring to
a definite map (denoted $q$ in this paper) and should not be confused with
the introduction of Planck's constant or second quantization.
Since any Dirac operator is assigned a negative parity, it remains unclear
how one could possibly include a ``mass term'' in its definition. 
No doubt, ``masses'' in physics are thought of as certain constants having 
some physical dimension but no parity whatsoever. Fortunately, the theory of 
superconnections comes to our rescue. The reason is, Dirac operators 
associated to a superconnection naturally include, apart from other terms, 
0-forms, formally sections of some endomorphism bundle. There is no problem of
assigning a negative parity to a 0-form. Physicists discovered the 0-form
and its role long time ago and named it {\em the Higgs field}. While the
gauge potential can only be defined on local patches of the manifold, the
Higgs field is a global object.

The rigidity of the mathematical setup forces us to
accept the following consequence. Nonzero masses of fermions can only
be created through the Higgs mechanism, i.e., by a spontaneous breakdown of
symmetry. Moreover, the Yukawa interaction ought to be regarded as a close
companion of the interaction between fermions and the gauge field, whatever
the gauge group.

Dirac operators as powerful tools in global analysis and index theory have a 
long history with constantly growing sophistication. But it is only recent 
that we have witnessed the introduction and study of remarkable connections 
between the abstract theory and the observable world. The modern theory of 
Dirac operators on spin manifolds began with the Atiyah-Singer Index Theorem 
[3]. A standard reference for spin geometry is the book by Lawson and Michelson
[4]. As we will see, there is a 1:1 correspondence between Dirac operators 
and Clifford superconnections. The theory of superconnections has been 
initiated by Quillen [5][6].

A comprehensive account of the role of Dirac operators in relativistic
quantum mechanics may be found in Thaller's book [7].
Dirac operators on twisted spinor modules, suitably chosen for the action
functional of the Standard Model, have been introduced by Tolksdorf [8].
Connes' work on non-commutative geometry [9] also puts Dirac operators into
the center of his approach and thus relates geometry to quantum theory [10].
The idea that quantum field theory provides a natural framework for problems
in differential topology and geometry has been pursued by Witten in a
number of papers (see [11] for instance). An attempt to combine concepts
from the works of both Connes and Witten and to consider supersymmetry as a
unifying principle was initiated by Fr\"ohlich et.al.\ [12].

\section{Prerequisites}

We need some preparations from geometry. Along with the sketch
we introduce our notation. Differential geometry starts with the algebra
of functions,
$$
      \Gamma:=C^\infty(M)\,,
$$ 
on some smooth manifold $M$ of dimension $n$ having no boundary. 
Derivations of the algebra
$\Gamma$ are called {\em vector fields\/} for which we use capital letters:
$$
        X,Y,Z,\ldots \in \V :=\G{TM}\,.
$$
As a notational convention, we always write $\G{\E}$ for the space of smooth 
sections of a vector bundle $\E$ on $M$. As physicists, we think of 
$\phi\in\G{\E}$ as some ``field'' carrying energy and momentum as well as 
specific quantum numbers such as `spin' and `charge'.

From their role as derivations, 
one infers that vector fields form an infinite-dimensional Lie algebra with 
product given by the commutator: $[X,Y]\in\V$. Elements of the dual are
called {\em differential 1-forms\/} for which we use small letters:
$$
        u,v,w,\ldots \in \V^*:=\G{T^*M}\,.
$$
For the canonical pairing we use brackets:
$$
         \br{\,,\,}: \V\times\V^*\to\Gamma\,.
$$
If the main concern is analysis on manifolds, one quickly passes to the
exterior algebra of differential forms:
$$
    \Om :=\G{\bw T^*M}=\sum_{p=0}^n\Om^p,\qquad \Om^p:=\G{\bw^p T^*M}\,.
$$
Though the algebra $\Om$ is $\ZZ$-graded, it may also be regarded 
$\ZZ_2$-graded: 
$$         
      \Om =\Om^+\oplus\Om^-, \qquad \Om^+
       =\sum_{p=\mbox{\scriptsize even}}\Om^p,\qquad
      \Om^-=\sum_{p=\mbox{\scriptsize odd}}\Om^p\,.
$$
The latter grading turns $\Om$ into a superalgebra with two consequences:
\begin{itemize}
\item $\Om$ is supercommutative, 
\item operators on $\Om$ are distinguished by their parity (even/odd).
\end{itemize}
Above all, one regards $\Om$ as a representation space for the 
following Lie superalgebra:
$$
\begin{tabular}[t]{lll}
$\sco{d,d}=0$       &                             &                         \\
$\sco{d,L_X}=0$     &$\sco{L_X,L_Y}=L_{[X,Y]}$    &                         \\
$\sco{d,\io(X)}=L_X$&$\sco{L_X,\io(Y)}=\io([X,Y])$&$\sco{\io(X),\io(Y)}=0$\,. 
\end{tabular}
$$
The interpretation of the expressions is as usual:
\begin{description}
\item[$d$\ \ \ \,:] exterior derivative. It has degree $+1$ (maps $\Om^p$
                    into $\Om^{p+1}$) and thus is of odd type. To state
                    that $\sco{d,d}=0$ is equivalent to saying that $d^2=0$.
\item[$L_X$\ :]     Lie derivative associated to a vector field $X$.
                    It has degree 0 and so is of even type. Its action
                    on 1-forms is described by
                    $$
                      \br{Y,L_Xv}=X\br{Y,v}-\br{[X,Y],v}\,.
                    $$
\item[$\io(X)$:]    contraction operator of degree $-1$ and 
                    hence of odd type. It contracts forms
                    $\phi\in\Om$ by the vector field $X\in\V$ so that
                    $$
                      \io(X)(v\wedge\phi)=\br{X,v}\phi-v\wedge\io(X)\phi\,.
                    $$
\end{description}
It proves convenient to add the {\em multiplication operator\/} $\ep(v)$ 
to this list. It is of degree $+1$ and hence of odd type. It multiplies forms 
$\phi\in\Om$ by $v\in\V^*$ from the left,
$$ 
                  \ep(v)\phi=v\wedge\phi\,,
$$
and satisfies the following algebraic relations:
$$
\begin{tabular}[t]{ll}
$\sco{\ep(u),\ep(v)}=0$      &$\sco{d,\ep(v)}=dv$ \\
$\sco{L_X,\ep(v)}=\ep(L_Xv)$ &$\sco{\io(X),\ep(v)}=\br{X,v}\,.$ 
\end{tabular}
$$
The supercommutator relations listed above have to be supplemented by
\begin{equation}
  \label{bas0}
     d1=0,\qquad L_X1=0,\qquad\io(X)1=0,\qquad\ep(v)1=v 
\end{equation}
($1\in\Gamma$). Since our prime interest lies in differential operators, 
we lay stress upon the structure of $\G{\E}$ as a $\Gamma$-module, no matter
what $\E$, i.e., global sections (or ``fields''), $\phi\in\G{\E}$, can be 
multiplied by functions $f,g\in\Gamma$ so that $(f+g)\phi=f\phi+g\phi$ and 
$(fg)\phi=f(g\phi)$. When dealing with superbundles, functions are viewed as 
operators of parity $+1$, acting on the left of $\G{\E}$. If $\E_1$ and
$\E_2$ are two vector bundles, we may form their tensor product $\E_1\otimes
\E_2$ and get an isomorphism
$$
       \G{\E_1}\otimes_\Gamma\G{\E_2}\cong \G{\E_1\otimes\E_2}.
$$
Thus, if $A_i$ are operators on $\G{\E_i}$, their tensor product, 
$A_1\otimes A_2$, is well defined only in situations where $[A_i,f]=0$ for
all $f\in\Gamma$. Operators possessing the latter property are said to
be {\em local operators\/} as opposed to differential operators. 

While the exterior derivative $d$ and the Lie derivative $L_X$ are differential
operators, the contraction operator $\io(X)$ and the multiplication operator
$\ep(v)$ are local:
$$
 \sco{d,f}=\ep(df),\qquad\sco{L_X,f}=Xf,\qquad 
 \sco{\io(X),f}=0, \qquad\sco{\ep(v),f}=0\,. 
$$
Suppose the manifold is orientable and $\om_0\in\G{\bw^nT^*M}$ is a volume
form. Then each of the relations (\ref{bas0}) finds a parallel in
\begin{equation}
   \label{basn}
      d\om_0=0,\qquad L_X\om_0=d\sigma_X,
       \qquad\io(X)\om_0=\sigma_X,\qquad \ep(v)\om_0=0
\end{equation}
where $\sigma_X\in\G{\bw^{n-1}T^*M}$ is the surface form in the direction
of $X$. If $X$ is compactly supported and $\mbox{supp}(X)\subset G$, then
by Stokes' Theorem
\begin{equation}
  \label{stokes}\textstyle
        \int_G d\sigma_X= \int_{\partial G} \sigma_X=0
\end{equation}
since $\sigma_X$ vanishes on the boundary $\partial G$ (or $\partial G=
\emptyset$).

\section{Riemannian Manifolds \\ and Levi-Civita Connection}

A non-degenerate symmetric bilinear form in each tangent space 
$T_xM$, which depends smoothly on $x\in M$, defines a semi-Riemannian geometry.
If in addition the bilinear form is positive definite (i.e., a scalar
product), we are dealing with a Riemannian manifold in the traditional sense.
Therefore, a (semi-)Riemannian manifold induces a (pseudo-)Euclidean structure
on each tangent space. In essence, the semi-Riemannian geometry provides a 
bilinear form $(X,Y)\in\Gamma$ in $\V$. 

Locally (on a chart), one can choose a basis $(\pd_i)_{i=1}^n$ so that
$[\pd_i,\pd_k]=0$. Any vector field assumes the form $X=v^i\pd_i$
and the metric tensor is given by
$$
   g=(g_{ij}),\qquad g_{ij}:=(\pd_i,\pd_j)\qquad\quad i,j=1,\ldots,n\,.
$$

As the bilinear form $(,)$ on $\V$ is non-degenerate, there is a natural 
isomorphism $\V\to\V^*$ taking $Y$ into $Y^\#$ so that $\br{X,Y^\#}=(X,Y)$. 
Moreover, there is an induced symmetric bilinear form on 1-forms determined 
by the relation $(X^\#,Y^\#)=(X,Y)$. Locally, one 
may introduce the dual basis $dx^i$ in $\V^*$, satisfying $\br{\pd_i,dx^k}=
\delta_i^k$, so as to obtain $\pd_i^\#=g_{ij}dx^j$. 
The inverse of the metric tensor may be written
$$
     g^{-1}=(g^{ij}),\qquad g^{ij}:=(dx^i,dx^j)\qquad\quad i,j=1,\ldots,n\,.
$$
If $\alpha\in\Om$, then in local coordinates we would write $\alpha=\alpha_I 
dx^I$ where the coefficients $\alpha_I$ are local functions on $M$, the
sum is over all subsets $I\subset\{1,\cdots,n\}$, and
$$
    dx^I=dx^{i_1}\wedge\cdots\wedge dx^{i_p}\in\Om^p,\qquad i_1<\cdots <i_p
$$
($dx^\emptyset=1$) where $I=\{i_1,\cdots,i_p\}$ so that $p=|I|$.

A vector field $X$ is said to be the {\em gradient\/} of $f\in\Gamma$ and
one writes $X=\gr f$ provided $X^\#=df$. In local coordinates,
$$
             \gr f=g^{ij}(\pd_if)\pd_j\,.
$$
We will use the fact that gradients form a total set in $\V$ in the sense that
any vector field $X$ can be written as $\sum g_i\gr f_i$ for suitable functions
$f_i,g_i\in\Gamma$. Equivalently, $X^\#=\sum g_idf_i$.

The {\em divergence\/} of a vector field $X$ is conveniently defined by 
applying the Lie derivative $L_X$ to the volume form $\om_0$:
\begin{equation}
  \label{lx}
          L_X\om_0 \equiv d\sigma_X=(\DIV X)\om_0\,,
\end{equation}
Locally, if $X=a^i\pd_i$,
\begin{eqnarray}
  \label{dvx}
  \DIV X &=&\dgm\pd_i(a^i\dg)\\
         &=&\pd_ia^i+a^i\pd_i\log\dg\,.
\end{eqnarray}
The {\em Levi-Civita connection\/} $\nabla$, also called a {\em covariant 
derivative on the tangent bundle}, has two defining properties: 1.\ it is 
torsion-free,
$$
       \nabla_XY-\nabla_YX-[X,Y]=0\,,
$$
and 2.\ it preserves the Riemannian metric: 
\begin{equation}
  \label{riem}
    d(X,Y)=(\nabla X,Y)+(X,\nabla Y)\ \in \V^*\,.  
\end{equation}
One interprets $\nabla$ as a map from $\V$ to $\G{T^*M\otimes TM}$,
while $\nabla_X$ is a differential operator on $\V$ obtained from $\nabla$ by
contraction: $\nabla_X=\io(X)\circ\nabla$.
The local relation, which corresponds to Eq.(\ref{riem}),
$$
     \nabla\pd_j=\Gamma_{ij}^k\, dx^i\otimes\pd_k
$$
introduces the Christoffel symbols. In terms of the metric tensor,
$$
\Gamma_{ij}^k=\hf g^{kl}(\pd_ig_{jl}+\pd_jg_{il}-\pd_lg_{ij})\,.
$$
Duality carries the Levi-Civita connection from the tangent bundle to the 
cotangent bundle (with same notation):
$$
     d\br{X,v}=\br{\nabla X,v}+\br{X,\nabla v}\ \in\V^*\,.
$$
Therefore, $\nabla$ may also be interpreted as a map from
$\V^*$ to $\G{T^*M\otimes T^*M}$ such that 
\[
   d(u,v)=(\nabla u,v)+(v,\nabla v)\ \in\V^*\,.
\]
A routine calculation shows that, locally, we have
\begin{equation}
    \nabla dx^j=-\Gamma_{ik}^j\,dx^i\otimes dx^k\,.\label{local}
\end{equation}
Note that the Christoffel symbols enter here with a minus sign. 

Contractions $\iota$ appear under various circumstances. One is provided by the
isomorphism $\V\cong\V^*$. Whenever $X^\#=u$, we write $\io(u)$ in place 
of $\io(X)$ so that 
$$
           \sco{\io(u),\ep(v)}=(u,v)\qquad (u,v\in\V^*)\,.
$$
Another occurrence is the trace
$$
   \G{T^*M\otimes T^*M}\arr{\io}\Gamma,\qquad \io(u\otimes v)=(u,v)\,,
$$
or equivalently,
$$
   \G{T^*M\otimes TM}\arr{\io}\Gamma,\qquad \io(u\otimes X)=\br{X,u}\,.
$$
Still another occurrence is the partial trace
$$
   \G{T^*M\otimes\bw T^*M}\arr{\io}\Om,\qquad \io(u\otimes\phi)=\io(u)\phi
$$
where $\phi\in\Om$. All these different notions of `contraction'
will play a role, in some way or another, in what follows.

In the next step, $\nabla_X$ is extended to some differential 
operator on $\Om$ satisfying 
$$
    \sco{\nabla_X,\io(v)}=\io(\nabla_Xv),\qquad 
    \sco{\nabla_X,\ep(v)}=\ep(\nabla_Xv)\,.
$$
This shows that the action on 0/1-forms is the only piece of information we 
need. While the action on 0-forms $f$ is given by the formula 
$\nabla_Xf=(X,df)=Xf$, the action on 1-forms $v$ is determined by 
$$
   X\br{Y,v}=\br{\nabla_XY,v}+\br{Y,\nabla_Xv}\,.
$$
Therefore, in local coordinates, $\nabla_{\pd_i}dx^j=-\Gamma_{ik}^jdx^k$.

\section{Generalized Laplace Operators}

In the sequel, $\E$ is some vector bundle based on a (semi-)Riemannian 
manifold $M$ and $\G{\E}$ is the $\Gamma$-module of sections.
It is very easy to provide a global characterization of what is meant by a 
first-order differential operator $D$ on the vector bundle $\E$:
$$
    \G{\E}\arr{D}\G{\E},\qquad [D,f]=c(df)\qquad (f\in\Gamma). 
$$
Here, $c:\G{T^*M}\to\G{\End\E}$ may be any $\Gamma$-linear map. For instance,
if $\E=TM$ or $\E=T^*M$ and $\nabla$ is the Levi-Civita connection,
$\nabla_X$ is such an operator with $c(df)=Xf$.

It is straightforward to write down similar conditions for second-order
differential operators. Among them we single out those operators $H$ that
could justly be called {\em generalized Laplacians\/}:
\begin{equation}
  \label{lap}
    \G{\E}\arr{H}\G{\E},\qquad [[H,f],g] +2(df,dg)=0\qquad(f,g\in\Gamma).
\end{equation}
On a local chart and to leading order, $H$ is of the form
$$
        H=-g^{ij}\pd_i\pd_j+\,\mbox{first and zero order terms}\,.
$$
Zero order terms in $H$ are frequently referred to as {\em local operators}.
In general, we lend this name to operators that commute with $f\in\Gamma$,
and, in the present context, they may be thought of as elements of
$\G{\End\E}$.

Until now there have been no restrictions on the first and zero order terms 
in a local decomposition of $H$. However, given a connection 
$$
  \nabla^\E:\Gamma(\E)\to\Gamma(T^*M\otimes \E),\qquad [\nabla^\E,f]=df
$$
on the bundle $\E$, one obtains a Laplacian which fixes these terms. 
Namely, combining $\nabla^\E$ with the Levi-Civita connection $\nabla$ 
on the cotangent bundle, one constructs, on the product bundle 
$\Ff=T^*M\otimes\E$, a connection\footnote{While products like
$\nabla\otimes\one$ and $\one\otimes\nabla^\E$ separately are meaningless, 
their sum nevertheless has a meaning owing to the consistency relation
$[\nabla,f]=[\nabla^\E,f]$ ($f\in\Gamma$) which guarantees that  
applying $\nabla^\Ff$ to $(fv)\otimes\phi$ and $v\otimes(f\phi)$ gives the
same result.} 
$$
          \nabla^\Ff :\G{\Ff}\to\G{T^*M\otimes\Ff}
$$
as $\nabla^\Ff=\nabla\otimes\one+\one\otimes\nabla^\E$ and in a second step, 
composing two maps, one gets
$$
   \G{\E}\arr{\nabla^\E}\G{T^*M\otimes\E}
    \arr{\nabla^\Ff}\G{T^*M\otimes T^*M\otimes\E}\,.
$$
Finally, one takes advantage of the trace on $T^*M\otimes T^*M$ (see the end
of Section 3):
$$
    \G{T^*M\otimes T^*M\otimes\E}\ \arr{\io}\ \G{\E}\,.
$$
The Laplacian associated with the connection $\nabla^\E$ is thus given by
\begin{equation}
    \triangle^\E=-\io\circ\nabla^{\Ff}\circ\nabla^\E\ :
   \ \Gamma(M,\E)\to\Gamma(M,\E)\,.\label{kLO}
\end{equation}
Readers, who wish to see what this construction means locally, should do the
following calculation. Assuming $\nabla^\E=dx^i\otimes(\pd_i+A_i)$ so that
$\nabla^\E_{\pd_i}=\pd_i+A_i$ where the $A_i$ are local sections of
the endomorphism bundle $\End\E$, they will find
\begin{eqnarray*}
  \triangle^\E&=&-\io(\nabla\otimes\one+\one\otimes 
                 dx^i\otimes(\pd_i+A_i))\circ dx^j\otimes (\pd_j+A_j)\\
              &=&-\io((\nabla dx^j)\otimes(\pd_j+A_j)+
                 dx^j\otimes dx^i\otimes(\pd_i+A_i)(\pd_j+A_j))\\
              &=&-\io(-\Gamma_{ik}^jdx^i\otimes dx^k\otimes(\pd_j+A_j)
                 +dx^j\otimes dx^i\otimes(\pd_i+A_i)(\pd_j+A_j))\\
              &=&-g^{ij}(\pd_i+A_i)(\pd_j+A_j)+g^{ik}\Gamma_{ik}^j(\pd_j+A_j)\\
              &=&-g^{ij}(\nabla^\E_{\pd_i}\nabla^\E_{\pd_j}-
                 \Gamma_{ij}^k\nabla^\E_{\pd_k})\,.
\end{eqnarray*}
We will frequently use the fact that, on a local chart, the connection
$\nabla^\E$ is determined by $dx^i\otimes A_i$, called the {\em connection
1-form\/} in geometry or the {\em vector potential\/} in physics.

We will now investigate the question whether every Laplacian is obtained this
way, i.e., whether it is associated to some connection on $\E$. Take
$f\in\Gamma$ and consider the following first order differential operator on 
$\G{\E}$:
\begin{eqnarray*}
[\triangle^\E,f]&=&-g^{ij}[\nabla^\E_{\pd_i}\nabla^\E_{\pd_j}-\Gamma^k_{ij}
                   \nabla^\E_{\pd_k},f]\\
                &=&-g^{ij}(\nabla^\E_{\pd_i}[\nabla^\E_{\pd_j},f]+
                  [\nabla^\E_{\pd_i},f]\nabla^\E_{\pd_j}
                  -\Gamma^k_{ij}[\nabla^\E_{\pd_k},f])\\
                &=&-g^{ij}(\nabla^\E_{\pd_i}(\pd_jf)+(\pd_if)\nabla^\E_{\pd_j}
                  -\Gamma^k_{ij}\pd_kf)\\
                &=&-g^{ij}(\pd_i\pd_jf
                   -\Gamma^k_{ij}\pd_kf+2(\pd_jf)\nabla^\E_{\pd_i})\\
                &=&-(\nabla,df)-2(df,\nabla^\E)\\
                &=&\triangle f-2\nabla^\E_X\,,\qquad X=\gr f\,.
\end{eqnarray*}
The result establishes a relationship with the Laplace-Beltrami operator
$\triangle$ acting on $f\in\Gamma$,
$$
   \triangle f= -(\nabla, df)= d^*df \ \in\Gamma\,,
$$
which, locally, is of the form 
$$ 
       \triangle f:=-g^{ij}(\pd_i\pd_jf-\Gamma_{ij}^k\pd_kf)\,.
$$ 
It should be clear that the covariant derivative $\nabla^\E_X$ is 
completely specified by giving its value for gradients $X=\gr f$ only,
since gradients form a total set in $\V$, and $X\mapsto\nabla^\E_X$ 
can always be extended by linearity. 

Thus, given an arbitrary Laplacian $H$ on the bundle $\E$, we may associate 
to it a connection $\nabla^\E$ on $\E$ such that
$$
  \nabla^\E_X =\hf(\triangle f-[H,f]),\qquad  X=\gr f\,.
$$
By the above construction and taking into account Eq.(\ref{lap}), we obtain
\begin{equation}
      \label{LO}
            [\nabla^\E_X,g]=-\hf[[H,f],g]=(df,dg) =\br{X,dg}
\end{equation}
equivalent to $[\nabla^\E,g]=dg$ which proves that $\nabla^\E$ is indeed
a connection on $\E$.
Now, for $f\in\Gamma$ and $X=\gr f$,
$$
         [H,f]=\triangle f-2\nabla^\E_X=[\triangle^\E,f]
$$
and hence $[H-\triangle^\E,f]=0$ which shows that the Laplacian we started
from has the general form
\begin{equation}
  \label{deco}
          H=\triangle^\E+F, \qquad F\in\G{\End\E}\,.  
\end{equation}
This decomposition of $H$ into a canonical Laplacian $\triangle^\E$ and a 
local operator $F$ is unique as can be inferred from the above discussion.

Thus, to summarize, we may state that a generalized Laplacian on a vector 
bundle $\E$ is completely characterized by giving three different data:
\begin{itemize}
\item a Riemannian metric on the manifold $M$,
\item a connection $\nabla^\E$ on the bundle $\E$,
\item a local operator $F\in\G{\End\E}$.
\end{itemize}

\section{Generalized Dirac Operators}

We will now be more specific and assume that $\E$ is a superbundle
(whose fibers are superspaces), i.e., we write $\E=\E^+\oplus\E^-$.
By a generalized Dirac operator $D$ on $\E$ we mean some first-order
differential operator of odd type (negative parity)
$$      
                D:\G{\E^\pm}\to\G{\E^\mp}\,,
$$
whose square $D^2$ is a generalized Laplacian. From now on we drop the
decoration ``generalized'' in reference to Laplacians and Dirac operators.

As $D$ is of first order, there is a map
\begin{equation} 
  \G{T^*M}\arr{c}\G{\End^-\E},\qquad c(df)=[D,f]\qquad (f\in\Gamma)\,.
      \label{cact}
\end{equation}
Sometimes, a different, however equivalent interpretation of the same map 
seems more appropriate:
$$
  \G{T^*M\otimes\E^\pm}\arr{c}\G{\E^\mp},\qquad c(df\otimes\phi)=[D,f]\phi\,.
$$
Since $\E$ is a $\Gamma$-module, we have the inclusion 
$\Gamma\subset\G{\End^+\E}$. Knowing that $f$ (as an operator) has positive
parity, we may in Eq.(\ref{cact}) replace the commutator by the
the supercommutator, $c(df)=\sco{D,f}$, and form the square:
\begin{eqnarray*}
c(df)^2&=&\hf\sco{ c(df),c(df)} \ =\ \hf\sco{ \sco{ D,f},\sco{ D,f}}\\
       &=&\hf(\sco{ D,\sco{ f,\sco{ D,f}}}-\sco{ f,\sco{ D,\sco{ D,f}}})\\
       &=&-\hf\sco{ f,\sco{ D,\sco{ D,f}}}\\
       &=&\frc{1}{4}(\sco{\sco{D,\sco{ D,f}},f}+\sco{\sco{D,\sco{D,f}},f})\\
       &=&\frc{1}{4}\sco{\sco{\sco{ D,D},f},f}\ =\ \hf\sco{\sco{ D^2,f},f}\,.
\end{eqnarray*}
Here we have used two facts: first, the operators $c(df)$ and $D$ are of odd 
type (lines 1 and 5) and second, the operator $\sco{D,f}$ is local (from line 
2 to 3). We have also used the Jacobi identity for supercommutators two times. 
The result may be written using ordinary brackets:
$$
         c(df)^2=\hf[[D^2,f],f]\,.
$$
As yet we have not exploited the condition that $D^2$ is a  
Laplacian. If we do, our result becomes even more simple:
$$
          c(df)^2+(df,df)=0\,.
$$
We note that $(df,df)$ induces a quadratic form on each cotangent space
$T^*_xM$ and $c$ restricts to a Clifford map $T^*_xM\to\End_x\E$. Let
$C(M)$ denote the Clifford bundle over $M$, having the (real) Clifford algebras
$C(T^*_xM)$ as fibers. Owing to the universal property of Clifford algebras,
the Clifford map $c$ can uniquely be extended to an algebraic homomorphism
$$
         \G{C(M)}\arr{c}\G{\End\E}\,.
$$
called the Clifford action on the bundle $\E$. An alternative
description of the same map is
$$
       \G{C(M)\otimes\E}\arr{c}\G{\E},\qquad c(a\otimes\phi)=c(a)\phi
$$
where $a\in\G{C(M)}$ and $\phi\in\G{\E}$. To summarize:
\begin{quote}
\em
A first-order differential operator $D:\G{\E^\pm}\to\G{\E^\mp}$
is a Dirac operator iff the map $c:\G{T^*M}\to\G{\End^-\E}$ 
given by $c(df)=\sco{D,f}$ defines a Clifford action on $\E$, i.e., turns
the bundle $\E$ into a Clifford supermodul.
\end{quote}

\section{Connections and Dirac Operators}

The way Dirac operators enter physical theories is through connections.
Their mutual relationship will be the subject of this section. 
A connection $\nabla^\E$ on the superbundle $\E$ will always be 
assumed to respect the grading:
$$
             \nabla^\E:\G{T^*M\otimes\E^\pm}\to\G{\E^\pm}\,.
$$
Locally, a connection (also called a covariant derivative) may be written
$$
    \nabla^\E=dx^i\otimes\nabla_{\pd_i}^\E=dx^i\otimes(\partial_i+A_i)
$$
where the $A_i$ are local sections of the endomorphism bundle $\End^+\E$.
In physical terms, they are the components of the vector potential.

Any connection gives rise to a canonical Dirac operator $\nsl^\E$
through a composition of maps:
$$
  \nsl^\E:\G{\E^\pm}\arr{\nabla^\E}\G{T^*M\otimes\E^\pm}\arr{c}\G{\E^\mp}\,.
$$
Locally, it reduces to an expression familiar from Dirac's theory:
$$
  \nsl^\E=c(dx^i\otimes\nabla^\E_{\pd_i})
  =c(dx^i)\nabla^\E_{\pd_i}=c(dx^i)(\pd_i+A_i)\equiv\psl+\asl\,.
$$
Correctness of the definition is checked immediately:
$$
    \sco{\nsl^\E,f}=[c(dx^i)\nabla^\E_{\pd_i},f]
    =c(dx^i)[\nabla^\E_{\pd_i},f]=c(dx^i)\pd_if=c(df)\,.
$$
We will now focus on the set of connections in relation to the set
of Dirac operators. The first thing to be aware of is: the set of connections
$\nabla^\E$ on a superbundle $\E$ is an {\em affine space} modelled on the 
vector space of 1-forms, $\G{T^*M\otimes\End^+\E}$. That is to say, 
two connections differ by some element of that vector space. 
By the same token, Dirac operators with same Clifford map $c$ form an 
affine space. For, if
$D_1$ and $D_2$ are two Dirac operators, then $[D_1-D_2,f]=c(df)-c(df)=0$ and 
hence $D_1-D_2\in\G{\End^-\E}$. Two connections giving rise to the same
Dirac operator (with same Clifford map) will be termed {\em equivalent}.
Equivalence $\nabla_1^\E\sim\nabla_2^\E$ means 
$$ 
 \nabla_1^\E-\nabla_2^\E=A\in\G{T^*M\otimes\End^+\E},\qquad \asl\equiv c(A)=0\,.
$$
So the question arises: can we characterize those 1-forms $A$ which are
mapped to zero by the Clifford map? A related problem is to characterize
the set $K$ of the exact sequence
$$
   0\arr{\ }K\arr{\ }\G{T^*M\otimes\E}\arr{c}\G{\E}\arr{\ }0\,.
$$
The following procedure is due to Tolksdorf [8]. To get a handle on $K=\Ker c$,
note that the Riemannian metric is but a
section $g$ of the product bundle $T^*M\otimes T^*M$ which, in local
coordinates, may be written $g=g_{ij}dx^i\otimes dx^j$. Combining $g$
and the unit $\one\in\G{\End\E}$, one obtains a canonical covariant 2-tensor 
$$
            g\otimes\one\in\G{T^*M\otimes T^*M\otimes\End\E}\,.
$$
The application of the Clifford map $c$ to 2-tensors is a two-step process,
$$
  \G{T^*M\otimes T^*M\otimes\End\E}\arr{c}
  \G{T^*M\otimes\End\E}\arr{c}\G{\End\E}\,,
$$
where it is understood that
$$
  \alpha\otimes\beta\otimes a\ \mapsto\ \alpha\otimes c(\beta)a
 \ \mapsto\ c(\alpha)c(\beta)a\,.
$$
Application to the canonical 2-tensor defines the 1-form $\om$ as the
middle term:
$$
     g\otimes\one\ \mapsto\ \om\ \mapsto\ -n\one\,.    
$$
The latter assertion may be checked locally:
\begin{eqnarray*}
  c(\om)&=&g_{ij}c(dx^i)c(dx^j)\ =\ \hf g_{ij}\sco{c(dx^i),c(dx^j)}\\
          &=&-g_{ij}(dx^i,dx^j)\one\ =\ -g_{ij}g^{ij}\one\ =\ -n\one\,.
\end{eqnarray*}
Thus the renormalized map
$$
     \G{\E}\arr{b}\G{T^*M\otimes\E},\qquad b(\phi)=-n^{-1}\om\phi     
$$
is a right-inverse of the map $\G{T^*M\otimes\E}\arr{c}\G{\E}$. Namely we have
$$
        c\circ b=\one,\qquad b\circ c =p\,.
$$
and $p$ must be a projector. Consequently, $K=\Ker p=\Ima(\one-p)$. 
Obviously, $K\subset\Ker p$ directly follows from $p=b\circ c$ 
while $\ker p\subset K$ follows from $c\circ p=c\circ b\circ c=c$.
The representation of $K$ as image of the projector $\one-p$ makes
$K$ a calculable object.

Another virtue of the canonical map $b$ is that it allows to associate,
to any connection $\nabla^\E$ which fails to satify $c(\nabla^\E)=D$, another 
connection $\nabla^\E_b$ which does satisfy $c(\nabla^\E_b)=D$:
$$
   \nabla^\E_b = (\one-p)\circ\nabla^\E+b\circ D\,.
$$ 
A simple calculation reveals that $\nabla^\E_b$ is indeed a connection:
\begin{eqnarray*}
 [\nabla^\E_b,f]
 &=& (\one-p)([\nabla^\E,f])+b([D,f])\\
 &=& (\one-p)(df)+b(c(df))\\
 &=& (\one-p)(df)+p(df) \ =\ df\qquad(f\in\Gamma).
\end{eqnarray*}
Moreover, the relations $c\circ(\one-p)=0$ and $c\circ b=\one$ imply that 
$\nsl_b^\E=D$. Note also that $\nabla_b^\E=\nabla^\E+A$ where the 1-form
$A$ is obtained from the difference of two Dirac operators:
$$
       A=b\circ(D-\nsl^E)\in\G{T^*M\otimes\End^+\E}\,.
$$
At this point we wish to emphasize that the Dirac operators obtained from 
connection $\nabla^\E$ are of
very special type and, in order to have a better grasp on Dirac operators
in general, we need to extend the notion of a connection on a superbundle.

\section{Superconnections and Dirac Operators}

The key to a wider concept of what should be called a {\em connection\/}
is the extension of ``ordinary'' connections $\nabla^\E$ to larger objects.
Two new objects come into sight. First, there is the 
superspace of $\E$-valued differential forms,
$$
        \Om(\E)=\G{\bw T^*M\skt\E}=\Om^+(\E)\oplus \Om^-(\E)
$$
for which we have the obvious imbeddings
\begin{eqnarray*}
                 \G{\E} &\equiv& \G{\bw^0T^*M\otimes\E}\subset \Om(\E)\\
      \G{T^*M\otimes\E} &\equiv& \G{\bw^1T^*M\otimes\E}\subset \Om(\E)\,.
\end{eqnarray*}
We stress that $\Om(\E)$ is $\ZZ_2$-graded even if $\E$ is not. Third, there
is the superalgebra of differential forms taking values in the
endomorphism bundle $\End\E$:
$$
    \Om(\End\E)=\G{\bw T^*M\skt\End\E}=\Om^+(\End\E)\oplus \Om^-(\End\E)\,.
$$
Note that $\Om(\End\E)$ consists of operators acting on the superspace 
$\Om(\E)$. In what follows, we shall make frequent use of the 
natural embeddings
$$
        \Gamma\subset\Om\subset \Om(\End\E)
$$
and, therefore, $\Om$ acts on $\Om(\E)$ from the left:
$$
           \alpha(\beta\otimes\phi)=(\alpha\wedge\beta)\otimes\phi,\qquad
           \alpha,\beta\in\Om,\ \phi\in\G{\E}\,. 
$$
The action of $\Om$ is universal (``defined in the same way'') for all
vector bundles $\E$. It is thus natural to regard $\Om(\E)$ as an $\Om$-modul.
In the same way, we regard $\Om(\End\E)$ as an $\Om$-modul.

In essence, what we suggest here is to replace all $\Gamma$-modules 
previously considered
by their corresponding $\Om$-modules. Putting $\Om$-modules now in a prominent
place is motivated by the following observation: any connection $\nabla^\E$ on
the bundle $\E$ uniquely extends to a first-order differential operator of
degree $+1$ on $\Om(\E)$ (mapping $\Om^p(\E)$ into $\Om^{p+1}(\E)$):
$$
   \nabla^\E:\Om(\E)\to \Om(\E),\qquad 
   \sco{\nabla^\E,\alpha}=d\alpha,\qquad \alpha\in\Om\,.
$$
In particular, $\nabla^\E$ maps $\Om^\pm(\E)$ into $\Om^\mp(\E)$ and thus
is of odd type. 

As a natural generalization, we let a {\em superconnection\/} be any 
differential operator
\begin{equation}
  \label{sco}
   \D:\Om^\pm(\E)\to \Om^\mp(\E),\qquad 
   \sco{\D,\alpha}=d\alpha,\qquad \alpha\in\Om\,.
\end{equation}
By definition, $\D$ is of first order and odd type.
A superconnection is fixed by providing the following restricted maps:
$$
      \D_p:\G{\E}\to\G{\bw^pT^*M\otimes\E},\qquad p=0,\ldots,n.
$$
Since
\begin{equation}
         \sco{\D_p,f}=\cases{df & if $p=1$\cr 0 & otherwise\cr}
         \qquad(f\in\Gamma)\,,
   \label{SC}
\end{equation}
it is only one component, $\D_1$, that acts by differentiation 
and defines an ``ordinary'' connection
on the bundle $\E$ while all other contributions $\D_p$ ($p\ne 1$) are local
terms, i.e., may be lifted to operators $\D_p\in \Om^-(\End\E)$. In particular,
$\D_0:\G{\E^\pm}\to\G{\E^\mp}$. Stated differently, a superconnection admits
a decomposition
$$          
              \D=\nabla^\E+L,\qquad L\in \Om^-(\End\E)
$$
with $\nabla^\E$ some connection on the bundle $\E$.
On coordinate charts, we find
\begin{equation}
  \label{dsl}
     \D=\om+dx^i\otimes(\pd_i+\om_i)+
     \hf(dx^i\wedge dx^j)\otimes\om_{ij}+\ldots
\end{equation}
where the coefficients $\om_{i_1\ldots i_k}$ ($k=0,\ldots,n$) are local 
sections of $\End\E$ with parity $(-1)^{k+1}$ and antisymmetric under 
permutations of their indices.

Once more, we look for extensions. A superconnection $\D$ on $\E$ can
be carried to the endomorphism bundle $\End\E$ setting
$$
  \D \Phi:=\sco{\D,\Phi}\in\Om(\End\E)\,,\qquad \Phi\in\G{\End\E}\,.
$$
The claim that the resulting operator is local can be verified:
\begin{eqnarray*}
[\D \Phi,f]&=&\sco{\sco{\D,\Phi},f}\qquad\qquad (f\in\Gamma)\\
        &=&\sco{\D,\sco{\Phi,f}}-\pa{\Phi}\sco{\Phi,\sco{\D,f}}\\
        &=&-\pa{\Phi}\sco{\Phi,df}\ =\ \sco{ df,\Phi}\\
        &=&(df\otimes\one)(\one\otimes \Phi)
           -\pa{\Phi}(\one\otimes \Phi)(df\otimes\one)\\
        &=&df\otimes \Phi-df\otimes \Phi=0\,.
\end{eqnarray*}
We have used the Jacobi identity for 
supercommutators, $\sco{\D,f}=df$, and $\sco{\Phi,f}=0$.

To obtain Dirac operators from superconnections we appeal to the theory of
Clifford algebras [1] and in particular to the symbol map $\sigma$
which establishes an linear isomorphism between the Clifford algebra and the
exterior algebra over the same vector space. In terms of bundles,
$$
           C(M) \arr{\sigma} \bw T^*M
$$
or else, in terms of sections,
$$
    \G{C(M)\skt\E}\arr{\sigma}\Om(\E)=\G{\bw T^*M\skt\E}\,.
$$
The Dirac operator $\Dsl$ associated to the superconnection $\D$ is given
by concatenating three maps:
$$
 \Dsl:\G{\E}\arr{\D}\Om(\E)\arr{\sigma^{-1}}\G{C(M)\skt\E}\arr{c}\G{\E}\,.
$$
In the sequel, we shall refer to $q=c\circ\sigma^{-1}$ as the
{\em quantization map}. We must check whether
$\Dsl$ satisfies the condition required for a Dirac operator: 
$$
 \sco{\Dsl,f}=\sco{q(\D),f}=q(\sco{\D,f})=q(df)=c(df)\,.
$$
We have used Eq.(\ref{SC}) and the fact that $\sigma$ restricts to the
identity on $T^*M$.

Physicists who feel uneasy about the degree of abstractness present in
the above construction of $\Dsl$ are advised to resort to local coordinates
and introduce ``gamma matrices'' setting $\gamma^i=c(dx^i)$ so that
$$
   q(dx^{i_1}\wedge\cdots\wedge dx^{i_k})=
   \frac{1}{k!}\sum_\pi\mbox{sign}(\pi)\gamma^{i_{\pi(1)}}
   \cdots\gamma^{i_{\pi(k)}}
$$
($k=2,\ldots,n$) to get from Eq.(\ref{dsl})
$$
 \Dsl =\om +\gamma^i(\pd_i+\om_i)
       +\frc{1}{4}(\gamma^i\gamma^j-\gamma^j\gamma^i)\om_{ij}+\ldots
$$
One remark is in order. We often regard the basis $(dx^i)_{i=1}^n$ as a 
generating set, not only for the algebra $\Om$ but also for the algebra
$\G{C(M)}$. While $dx^i\wedge dx^j$ is the exterior product of two such
elements, $dx^idx^j$ is their Clifford product. The symbol map $\sigma$
establishes a relationship:
\begin{eqnarray*}
     \sigma(dx^idx^j) &=& dx^i\wedge dx^j-g^{ij}\\
     \sigma(dx^idx^jdx^k)&=& dx^i\wedge dx^j\wedge dx^k-g^{ij}dx^k
     +g^{ik}dx^j-g^{jk}dx^i\qquad\mbox{etc.}
\end{eqnarray*}
In particular, $\sco{dx^i,dx^j}=-2g^{ij}$.

\section{Vector Fields Revisited}

This section should be regarded a footnote to the previous discussion.
We state some supplementary results, but skip the proofs.

When studying the exterior derivative $d$ on vector fields,
one introduces the exterior bundle $\bw TM$, the space of sections,
$\G{\bw TM}$, and defines, for arbitrary $X_i\in\V$,
$$
  d(X_1\wedge X_2\wedge\cdots\wedge X_p)=
  \sum_{i<j}(-1)^{i+j-1}[X_i,X_j]\wedge X_1\wedge\cdots
  \hat{X}_i\wedge\ldots\wedge\hat{X}_j\wedge\ldots\wedge X_p    
$$
where the hat $\hat{\phantom{a}}$ means elimination. This shows that the
exterior derivative has degree $-1$ on $\G{\bw TM}$, i.e., it maps
$\G{\bw^pTM}$ into $\G{\bw^{p-1}TM}$. There are obvious links to the
Cartan-Eilenberg (co)homology theory of Lie algebras.

The canonical pairing $\V\times\V^*\to\Gamma$ extends to a pairing
$$
          \br{\cdot,\cdot}\ :\ \G{\bw^p TM}\times\G{\bw^p T^*M}\to\Gamma\,,
          \qquad p=0,\ldots,n
$$
and one finds, for $\alpha\in\Om^p=\G{\bw^p T^*M}$,
\begin{eqnarray}
  \label{pair1}
   \br{X_0\wedge\ldots\wedge X_p,d\alpha} &+&
   \br{d(X_0\wedge\ldots\wedge X_p),\alpha}\nonumber\\
   &=&\sum_{i=0}^p (-1)^i X_i
   \br{X_0\wedge\ldots\wedge\hat{X}_i\wedge\ldots\wedge X_p,\alpha}\,. \qquad 
\end{eqnarray}
This formula can be generalized to cover situations where one deals with
a connection $\nabla^\E$ on a vector bundle $\E$. For this purpose,
the bracket receives a different but obvious interpretation:
$$
     \br{\cdot,\cdot}\ :\ \G{\bw^p TM}\times\G{\bw^p T^*M\otimes\E}\to\G{\E},
     \qquad p=0,\ldots,n.
$$
The fundamental formula generalizing (\ref{pair1}) says that, for 
$\Phi\in\Om^p(\E)$,
\begin{eqnarray}
  \label{pair2}
   \br{X_0\wedge\ldots\wedge X_p,\nabla^\E\Phi}&+&
   \br{d(X_0\wedge\ldots\wedge X_p),\Phi}\nonumber\\ 
   &=&\sum_{i=0}^p (-1)^i \nabla_{X_i}^\E
   \br{X_0\wedge\ldots\wedge\hat{X}_i\wedge\ldots\wedge X_p,\Phi}\qquad  
\end{eqnarray}
and demonstrates in a direct manner that the connection $\nabla^\E$ is
completely characterized by the action 
$$
          \nabla_X^\E :\G{\E}\to\G{\E}
$$
for every vector field $X$. The exterior derivative is but a special
case: $d$ is the unique connection of the trivial bundle $M\times\RR$ 
with $\Gamma$ its space of sections and
$$
             d_X:\Gamma\to\Gamma, \quad d_Xf=Xf\,.
$$
The question whether one can translate the content of (\ref{pair2})
into an algebraic language has an affirmative answer and leads to the
following characterization.
\begin{quote}\em
A connection on the vector bundle $\E$ is a first order differential
operator $\nabla^\E$ on $\Om(\E)$ of degree $+1$ satisfying
\begin{equation}
  \label{dsc}
  \sco{\sco{\nabla^\E,\io(X)},\io(Y)}=\io([X,Y])
\end{equation}
for arbitrary vector fields $X,Y$.
\end{quote}
Note that $[X,Y]=d(X\wedge Y)$. The proof that the Eqs.\ (\ref{pair2}) and 
(\ref{dsc}) are but different versions of the same condition relies on 
two formulas:
\begin{enumerate}
\item \qquad $\br{X\wedge A,\Phi}=\br{A,\io(X)\Phi},
      \qquad A\in\G{\bw TM},\ \Phi\in\Om(\E)$
\item \qquad $\nabla_X^\E\phi=\sco{\nabla^\E,\io(X)}\phi,
      \qquad \qquad \phi\in\G{\E}$
\end{enumerate}
It is easy to reproduce the result that two connections differ by a
1-form $A\in\G{T^*M\otimes\End^+\E}$ since $A$ must be a local operator
of degree $+1$ satisfying $\sco{\sco{A,\io(X)},\io(Y)}=0$.

Suppose $\E$ is a superbundle. Can we characterize superconnections $\D$ on 
$\E$ in the same manner, i.e., by imposing the condition (\ref{dsc}), but 
replacing the restriction degree$(\nabla^\E)=+1$ by parity$(\D)=+1$?
Generally, the answer is no. Reason: with $\D=\sum_{p=0}^n\D_p$ the canonical 
decomposition, one gets
$$
      \sco{\sco{\D_p,\io(X)},\io(Y)}=\cases{0 & if $p=0$\cr
            \io([X,Y])& if $p=1$\cr}
$$
while, for $p\ge 2$, the contributions $\sco{\sco{\D_p,\io(X)},\io(Y)}$ 
do not vanish for all $X$ and $Y$, unless $\D_p=0$. Motivated by this
observation we call $\D$ a {\em special superconnection\/}
if $\D_p=0$ for $p\ge 2$. We have the following obvious characterization.
\begin{quote}\em
A superconnection $\D$ on the superbundle $\E$ is special provided
$$
   \sco{\sco{\D,\io(X)},\io(Y)}=\io([X,Y])\,.
$$ 
Equivalently, the associated Dirac operator $\Dsl=q(\D)$ satisfies
$$
   \sco{\sco{\Dsl,q(u)},q(v)}=\sco{q(du),q(v)}
$$
for all 1-forms $u,v\in\V^*$.
\end{quote}
Note that
$$
   \sco{q(du),q(v)}-\sco{q(u),q(dv)}+2q(df)=0,\qquad f=(u,v).
$$    
It is not at all clear whether conditions involving a double
supercommutator are ``natural'' or ``desirable'' in some sense. 
The interesting feature of such conditions is that the operator $\D$ 
then becomes part of a Lie superalgebra. As for applications in physics, 
these conditions may be important because unwanted tensor fields (with $p\ge2$)
are excluded as constituents of a special superconnection, yet
leaving room for the scalar Higgs field (with $p=0$).

\section{Dirac Operators and Clifford Connections}

It is perhaps worth a digression to discuss an extension of the Levi-Civita
connection $\nabla$, this time going from the cotangent bundle to the Clifford 
bundle in a direct way (avoiding the exterior algebra and the
quantization map). By `extension' we mean a commuting diagram,
$$
\def\normalbaslines{\baselineskip30pt\lineskip5pt\lineskiplimit5pt}
\matrix{%
 \G{T^*M}&\mapr{\nabla}&\G{T^*M\otimes T^*M}\cr
\mapd{\ }&             &\mapd{\ }           \cr
 \G{C(M)}&\mapr{\nabla}&\G{T^*M\otimes C(M)}\cr} 
$$
where the vertical arrows correspond to the injection $T^*M\to C(M)$.
The condition, which guarantees uniqueness of the extension, uses the fact
that $\G{C(M)}$ is a superalgebra:
$$
      \nabla_X(ab)=(\nabla_Xa)b+a(\nabla_Xb),\qquad a,b\in\G{C(M)}\,.
$$
Or, to state the same in an abbreviated form,
$$
    \sco{\nabla,a}=\nabla a,\qquad  \nabla 1=0\,.
$$
As a consequence, the (extended) Levi-Civita connection
respects the $\ZZ_2$-grading of the Clifford bundle:
$$
             \G{C(M)^\pm}\arr{\nabla}\G{T^*M\otimes C(M)^\pm}\,.
$$
As before, we let $\E$ be some Clifford supermodule. Recall that the affine
space of Dirac operators has $\G{\End^-\E}$ as its underlying vector space.
From the discussion [1] we can draw important information about the structure
of the endomorphism bundle:
\begin{equation}\label{cong1}
\def\normalbaslines{\baselineskip30pt\lineskip5pt\lineskiplimit5pt}
\matrix{%
 \G{C(M)\skt\End_{C(M)}\E}&\mapr{\sigma}&\G{\bw T^*M\skt\End_{C(M)}\E}\cr
        \mapd{c}          &             &       \mapd{\ }             \cr
       \G{\End\E}         &\mapr{\ }    &\G{\bw T^*M\skt\End\E}       \cr}
\end{equation}
We denote by $\End_{C(M)}\E$ the superbundle of endomorphisms that
supercommute with the action of $C(M)$. More precisely, 
$\varphi\in\G{\End_{C(M)}\E}$ means that $\sco{c(a),\varphi}=0$ for all 
$a\in\G{C(M)}$. 
It is easily checked that $\G{\End_{C(M)}\E}$ inherits the structure of a 
superalgebra from $\G{\End\E}$. So the left vertical arrow in the above 
diagram, given by
$$
         c(a\otimes\varphi)=c(a)\varphi\,,
$$
should be understood as an isomorphism between superalgebras. However,
the upper horizontal arrow in the above diagram, known as 
the {\em symbol map}, is merely an isomorphism between superspaces.
In (\ref{cong1}), we also indicated two obvious inclusions. From the
previous discussion we know that
$$
         \Om^-(\End\E) \equiv  \G{(\bw T^*M\skt\End\E)^-}
$$
is the vector space underlying the affine space of superconnections.
It is definitely too
large to yield a 1:1 correspondence between Dirac operators $D$ and 
superconnections $\D$. To find an appropriate affine subspace of
(super)connections, we impose further conditions on $\nabla^\E$ resp.\ $\D$
using the Levi-Civita connection $\nabla$ on the Clifford bundle. 
Namely, $\nabla^\E$ is said to be a {\em Clifford connection\/} if
\begin{equation}
  \sco{\nabla^\E,c(a)}=c(\nabla a),\qquad a\in\G{C(M)}\,.
  \label{CC}
\end{equation}
If $\nabla^\E_1$ and $\nabla^\E_2$ are two Clifford connections on $\E$,
then
$$
    \nabla^\E_1-\nabla^\E_2\in \Om^1(\End_{C(M)}^+\E)\,.
$$
By Schur's Lemma, $\E$ is irreducible iff $\End_{C(M)}^+\E$ consists of
scalars only, more precisely, iff it reduces to a real or complex line 
bundle $\ell$,
with the consequence that the difference $\nabla^\E_1-\nabla^\E_2$ is a
real- or complex-valued 1-form, i.e., some element of $\Om^1(\ell)$.
   
In the same spirit, $\D$ is said to be a {\em Clifford superconnection\/} if
\begin{equation}
  \sco{\D,c(a)}=c(\nabla a),\qquad a\in\G{C(M)}\,.
  \label{CSC}
\end{equation}
If $\D_1$ and $\D_2$ are two Clifford superconnections on $\E$, then
$$
   \D_1-\D_2\in\Om^-(\End_{C(M)}\E)\cong\G{\End^-\E}
$$
and $\D_1-\D_2\in\Om^-$ if $\E$ is irreducible. This demonstrates:
\begin{quote}
\em There is a 1:1 correspondence between Clifford superconnections 
    and Dirac operators. 
\end{quote}
This particularly implies that there is a 1:1 correspondence between Clifford
superconnections and equivalence classes $[\nabla^\E]$ of connections
giving rise to the same Dirac operator $D$. Such a class may or may not
contain a Clifford connection $\nabla^\E$. If it does, we call $D$
a {\em standard Dirac operator}. This is a reasonable way to single out
special cases which are important in the sequel.

\section{Superconnections and Curvature}

Recall that a superconnection $\D$ on the superbundle $\E$ is a first-order 
differential operator of odd type on the $\Om$-module $\Om(\E)$:
$$
   \D:\Om^\pm(\E)\to \Om^\mp(\E),\qquad\sco{\D,\alpha}=d\alpha\qquad
    \alpha\in\Om\,.
$$
The {\em curvature\/} is defined as the square of
the superconnection and so is no longer a differential but a local operator
of even type:
$$
          \F=\D^2=\hf\sco{\D,\D}\in \Om^+(\End\E)\,.
$$
Locality, i.e., $[\F,f]=0$, follows from $d^2=0$. Actually, we may prove more:
$$
   \sco{\F,\alpha}=\hf\sco{\sco{\D,\D},\alpha}=\sco{\D,\sco{\D,\alpha}}
   =d^2\alpha=0\qquad (\alpha\in\Om).
$$
Taking $\E=T^*M$ and $\D=\nabla$, the Levi-Civita connection, we see that
the {\em Riemannian curvature\/} $R$ appears as a special case:
$$
      R=\nabla^2\in \Om^+(\End T^*M)\,.
$$
Moreover, owing to the special nature of the Levi-Civita connection $\nabla$, 
the curvature $R$ is a 2-form. In local coordinates,
$$
      R=\hf(dx^i\wedge dx^j)\otimes R_{ij},\qquad
      R_{ij}=\sco{\nabla_{\pd_i},\nabla_{\pd_j}}
$$
where the $R_{ij}$ are local sections of endomorphism bundle 
$\End T^*M$. To obtain coefficients in $\Gamma$, we pass to the
Riemann tensor:
$$
   \br{\pd_i,R_{kl}dx^j}={{R_i}^j}_{kl}=\pd_l\Gamma^j_{ki}-\pd_k\Gamma^j_{li}
        +\Gamma^j_{lm}\Gamma^m_{ki}-\Gamma^j_{km}\Gamma^m_{li}\,.
$$
Contractions lead to the Ricci tensor and the scalar curvature:
$$
    \mbox{Ric}_{ij}={R^k}_{ikj},\qquad r_M={R^{ij}}_{ij}\,.
$$
In the previous section the Levi-Civita connection has been shown to extend 
to the Clifford bundle $C(M)$. We may go one step further and extend it to the
larger bundle $\bw T^*M\skt C(M)$ by way of a commuting diagram,
$$
\def\normalbaslines{\baselineskip30pt\lineskip5pt\lineskiplimit5pt}
\matrix{%
 \G{C(M)}&\mapr{\nabla}&\G{T^*M\otimes C(M)}\cr
\mapd{\ }&             &\mapd{\ }           \cr
 \Om(C(M))&\mapr{\nabla}&\Om(C(M))\cr}
$$
and the condition
$$
                \sco{\nabla,\alpha}=d\alpha\qquad (\alpha\in\Om)
$$
using the structure of $\Om(C(M))$ as an $\Om$-module.
Consequently, the curvature $R=\nabla^2$ extends to a 2-form with values
in the endomorphism bundle $\End^+ C(M)$ satisfying
$$
                  R(ab)=(Ra)b+a(Rb),\qquad a,b\in\G{C(M)}
$$
which suggests the Ansatz $Ra=\sco{\R,a}$ with $\R$ some 2-form
taking values in the bundle $C^+(M)$. Locally, we may thus write
$$
    \R=\hf(dx^i\wedge dx^j)\otimes\R_{ij}\,\qquad
    \R_{ij}\in \Gamma_{\rm loc}(C^+(M)).
$$
From $[\R_{ij},dx^k]=R_{ij}dx^k=R_{l\:ij}^{\:k}dx^l$ and the Clifford
relation $\sco{dx^i,dx^j}=-2g^{ij}$ we infer that
$$
    \R_{ij}:=-\frc{1}{4}R_{klij}dx^kdx^l\,.
$$
For $\E$ some Clifford superbundle with Clifford action $c$, we may carry the 
curvature 2-form to $\E$ setting $R^\E=c(\R)$. In local coordinates,
$$
    R^\E:=\hf(dx^i\wedge dx^j)\otimes c(\R_{ij})\,\qquad 
           c(\R_{ij})\in\Gamma_{\rm loc}(\End^+\E).
$$
The remarkable property to be noticed is
$$
    \sco{R^\E,c(a)}=\sco{c(\R),c(a)}=c(\sco{\R,a})=c(Ra),\qquad a\in\G{C(M)}
$$
which parallels a similar relation involving the curvature $\F=\D^2$ of the 
superconnection $\D$:
$$
   \sco{\F,c(a)}=\hf\sco{\sco{\D,\D},c(a)}
                =\sco{\D,\sco{\D,c(a)}}=c(\nabla^2a)=c(Ra)\,.
$$
Combining these two relations, we derive a decomposition theorem for the
curvature: 
$$
         \F=R^\E +F^{\rm tw}, \qquad \sco{F^{\rm tw},c(a)}=0
$$
The first term is ``Riemannian'' while the second term, the {\em twisting 
part}, is a 2-form with values in $\End_{C(M)}^+\E$.

\section{Standard Dirac Operators and Curvature}

Let us now turn to the Dirac operators and investigate their squares,
the Laplacians, in relation to the concept of curvature. As has been
previously shown, the Laplacian associated to a Dirac operator
splits into a `canonical part' and some local operator. It is the 
local part that we wish to study in detail.

Standard Dirac operators $D=\nsl^\E$ are special in that they derive from
Clifford connections $\nabla^\E$ (see Section 8) which facilitates
the calculation of the square.
Let us therefore assume that $\nabla^\E$ is some Clifford connection with
curvature $F^\E=(\nabla^\E)^2$ so that, in a local description,
$$
   F^\E=\hf(dx^i\wedge dx^j)\otimes F^\E_{ij},\qquad 
   F^\E_{ij}=\sco{\nabla^\E_{\pd_i},\nabla^\E_{\pd_j}}\in
   \Gamma_{\rm loc}(\End^+\E).
$$
On the other hand, we get an explicit expression for the square of
the Dirac operator from a direct calculation:
\begin{eqnarray*}
(\nsl^\E)^2%
&=&\hf \sco{\nsl^\E,\nsl^\E}
    \ =\ \hf \sco{ c(dx^i)\nabla^\E_{\pd_i},c(dx^j)\nabla^\E_{\pd_j}}\\
&=&\hf \sco{ c(dx^i)\nabla^\E_{\pd_i},c(dx^j)}\nabla^\E_{\pd_j}
   -\hf  c(dx^j)\sco{ c(dx^i)\nabla^\E_{\pd_i},\nabla^\E_{\pd_j}}\\
&=&\hf  c(dx^i)\sco{\nabla^\E_{\pd_i},c(dx^j)}\nabla^\E_{\pd_j}
  +\hf \sco{ c(dx^i),c(dx^j)}\nabla^\E_{\pd_i}\nabla^\E_{\pd_j}\\
& &-\hf  c(dx^j)c(dx^i)\sco{\nabla^\E_{\pd_i},\nabla^\E_{\pd_j}}
   -\hf  c(dx^j)\sco{ c(dx^i),\nabla^\E_{\pd_j}}\nabla^\E_{\pd_i}\\
&=&-g^{ij}\nabla^\E_{\pd_i}\nabla^\E_{\pd_j}
   +c(dx^i)\sco{\nabla^\E_{\pd_i},c(dx^j)}\nabla^\E_{\pd_j}
   +\hf c(dx^jdx^i)\sco{\nabla^\E_{\pd_j},\nabla^\E_{\pd_i}}\\
&=&-g^{ij}\nabla^\E_{\pd_i}\nabla^\E_{\pd_j}
   +c(dx^i)c(\nabla_{\pd_i}dx^j)\nabla^\E_{\pd_j}
   +\hf c(dx^jdx^i)F^\E_{ji}\\
&=&-g^{ij}\nabla^\E_{\pd_i}\nabla^\E_{\pd_j}
   -\Gamma^j_{ik}c(dx^idx^k)\nabla^\E_{\pd_j} +
   \hf q(dx^j\wedge dx^i-g^{ji})F^\E_{ji}\\
&=&-g^{ij}\nabla^\E_{\pd_i}\nabla^\E_{\pd_j}
   -\Gamma^j_{ik}\hf c(\sco{ dx^i,dx^k})\nabla^\E_{\pd_j} 
   +q(\hf(dx^i\wedge dx^j)\otimes F^\E_{ij})\\
&=&-g^{ij}(\nabla^\E_{\pd_i}\nabla^\E_{\pd_j}
   -\Gamma^k_{ij}\nabla^\E_{\pd_k})+q(F^\E)\\
&=&\triangle^\E+\Fsl^\E\,.
\end{eqnarray*}
We have used $\Gamma^j_{ik}=\Gamma^j_{ki}$, $F^\E_{ij}=-F^\E_{ji}$, and
the quantization map $q=c\circ\sigma^{-1}$.
As a result of this lengthy calculation, we learn that the local part of
the Laplacian is precisely the ``quantized'' version $q(F^\E)$
of the curvature $F^\E$ of the Clifford connection $\nabla^\E$.

From the previous section we know that the curvature $F^\E$ also admits a 
decomposition, i.e., $F^\E=R^\E +F^{\rm tw}$. We may thus proceed and
calculate the Riemannian part of $\Fsl^\E$:
\begin{eqnarray*}
\Rsl^\E&=&\hf q(dx^i\wedge dx^j)R^\E_{ij}
          =\hf c(dx^idx^j+g^{ij})R^\E_{ij}\\
       &=&\hf c(dx^idx^j)R^\E_{ij}\qquad\qquad
           (\mbox{since\ }R^\E_{ij}=-R^\E_{ji})\\
       &=&-\frc{1}{8}c(dx^idx^j)R_{klij}c(dx^kdx^l)\\
       &=&-\frc{1}{8}R_{ijkl}c(dx^idx^jdx^kdx^l)\,.
\end{eqnarray*}
In order to further simplify the result, we use the identity
$$
  dx^jdx^kdx^l=\sigma^{-1}(dx^j\wedge dx^k\wedge dx^l)
                -g^{jk}dx^l+g^{jl}dx^k-g^{kl}dx^j.
$$
Owing to the 1.\ Bianchi identity $\ep^{jkl}R_{ijkl}=0$, the contribution 
coming from the first term on the right, which is antisymmetric in $jkl$, 
vanishes and thus
\begin{eqnarray*}
  \Rsl^\E&=&\frc{1}{8}(R_{i\:jk}^{\:j}-R_{i\:kj}^{\:j}
            +R_{ik\:j}^{\:\:\:j})\,c(dx^idx^k)\\
         &=&-\frc{1}{4}R^j_{\:ijk}\,c(dx^idx^k)\qquad\qquad
            (\mbox{since\ }R_{ik\:j}^{\:\:\:j}=0)\\
         &=&-\frac{1}{4}\mbox{Ric\,}_{ik}\,c(dx^idx^k)\\
         &=&\frc{1}{4}r_M\qquad\qquad\qquad\qquad
           (\mbox{since\ }\mbox{Ric\,}_{ik}=\mbox{Ric\,}_{ki})\,.
\end{eqnarray*}
Putting pieces together, we get the {\em formula of Lichnerowicz\/} valid
for Clifford connections $\nabla^\E$:
\begin{equation}
    (\nsl^\E)^2=\triangle^\E+\frc{1}{4}r_M+\Fsl^{\rm tw}\,.
\label{Lichnerowicz}
\end{equation}

\section{Poincar\'e Duality, Hodge Operator, and Coderivative}

Once more we need some preparation from linear algebra. Consider a tripel
$$
                     V,\ (\cdot,\cdot),\ \om_0
$$
consisting of an $n$-dimensional real or complex vector space $V$ endowed with
a scalar product $(\cdot,\cdot)$ and a volume form $\om_0$ so that the 
structure group is either $SO(n)$ or $SU(n)$. We will use the fact that the 
scalar product extends to the exterior algebra $\bw V$ and that 
$\om_0\in\bw^nV$ with $(\om_0,\om_0)=1$. We will also consider the dual 
$(\bw V)^*=\bw V^*$ and the canonical pairing
$$
  \br{\cdot,\cdot}\ :\ \bw V\times\bw V^*\to\RR\mbox{ resp.\ }\CC\,.
$$
The existence of a scalar product provides a natural antilinear\footnote{%
In the case of real vector spaces, `antilinear' coincides with `linear'.
We adopt the convention that a scalar product in a complex vector space is
{\em antilinear\/} in the first of its two arguments.}
isomorphism $\bw V\to \bw V^*$, $v\mapsto v^\#$ so that $\br{v^\#,w}=(v,w)$.
By {\em Poincar\'e duality\/} we mean a linear isomorphism 
$P:\bw V\to\bw V^*$ taking $\bw^kV$ into $\bw^{n-k}V^*$ so that
$$
         \br{Pv,w}=(\om_0,v\wedge w),\qquad v,w\in \bw V\,.
$$
The combination of both isomorphisms -- one linear and one antilinear -- yields
an antilinear isomorphism known as the {\em Hodge operator\/}:
$$
    \bw V \arr{*}\bw V,\qquad (*v,w)=(\om_0,v\wedge w)\,.
$$
It is an isometry in the sense that $(v,w)=(*w,*v)$ (i.e., {\em antiunitary\/}
in the complex case). We discover a small defect though: the Hodge operator
is `almost involutive',
$$
          *(*v)=(-1)^{p(n-p)}v,\qquad v\in\bw^pV\,.
$$
If $n=$odd, however, this defect disappears giving $*(*v)=v$ for all 
$v\in\bw V$.

One obvious generalization immediately comes to our mind. The sequilinear
form $(,)$ need not really be a scalar product; it need only be nondegenerate.
This observation helps to encorporate pseudo-Euclidean vector spaces into the 
present investigation.
We may also carry the construction of the Hodge operator to all kinds of vector
bundles, real or complex, on oriented Riemannian or semi-Riemannian 
manifolds. For instance, on 
$$
          \Om=\sum_p\Om^p =\G{\bw T^*M},\qquad \Om^p=\G{\bw^pT^*M}
$$ 
we have the following explicit formula:
$$
  *(dx^{i_1}\wedge\ldots\wedge dx^{i_k})=\frc{1}{(n-k)!}\dg
    \ep^{i_1\ldots i_k}_{\qquad j_{k+1}\ldots 
     j_n}dx^{j_{k+1}}\wedge\ldots\wedge dx^{j_n}
$$
where $\ep_{1\ldots n}=1$, $\ep^{1\ldots n}=\det\,g^{-1}$. The volume form
may be written
$$
      \om_0=*(1)=\dg dx^1\wedge\ldots\wedge dx^n.
$$
The formal adjoint $d^*$ of the exterior derivative, called the {\em 
coderivative}, 
is defined by
\begin{equation}
  \label{CodDef}
   \int (d^*\alpha,\beta)\om_0 =\int (\alpha, d\beta)\om_0
\end{equation}
for compactly supported $\alpha,\beta\in\Om$. The coderivative has degree
$-1$, satisfies $d^{*2}=0$, and can explicitly 
be expressed with help of the Hodge operator:
\begin{equation} 
       d^*\alpha=(-1)^{n(p+1)+1}*d*\alpha,\qquad \alpha\in\Om^p\,.
  \label{Cod1}
\end{equation} 
The relationship reduces to $d^*=-*d*$ if $n=$even. It is a peculiar fact
that the dimension $n$ of the manifold $M$ enters the parity of the
Hodge operator: par$(*)=(-1)^n$. That is to say, the $*$ operator
either takes $\Om^\pm$ to $\Om^\pm$ ($n=$even) or takes $\Om^\pm$ to $\Om^\mp$
($n=$odd). This makes physics in even dimensions differ from physics in odd 
dimensions.
 
Recall now that $\Om$ is a representation space for the canonical 
anticommutation relation (CAR) of contraction and multiplication operators:
$$
       \sco{\io(u),\ep(v)}=(u,v)\in\Gamma,\qquad u,v\in\V^*\,.
$$
Our interest ultimately lies in the fact that $c=\ep-\io$ is a Clifford map 
and extends to the Clifford action on $\Om$ (see [1] for details).

From a different standpoint, we are dealing with maps
\begin{eqnarray*}
  \G{T^*M\otimes\bw T^*M}&\arr{\ep}&\Om,\qquad 
                                 \ep(v\otimes\alpha)=v\wedge\alpha\\
  \G{T^*M\otimes\bw T^*M}&\arr{\io}&\Om,\qquad
   \io(u\otimes(v\wedge\alpha))=(u,v)\alpha-v\wedge\io(u\otimes\alpha)         
\end{eqnarray*}
where $\alpha\in\Om$. With regard to the sequence
$$
    \Om\arr{\nabla}\G{T^*M\otimes\bw T^*M}\arr{\ep,\,\io}\Om\,,
$$
we claim the validity of two remarkable formulas:
\begin{equation}
       d=\ep\circ\nabla\,,\qquad d^*=-\io\circ\nabla\,.
  \label{Cod3}
\end{equation}
As we shall see, the proof needs the fact that the Levi-Civita connection is
torsionfree. As a consequence of (\ref{Cod3}), the differential operator
$$
            D=d+d^*=c\circ\nabla
$$ 
is a (standard) Dirac operator on $\Om$ whose square
$$
            D^2=\sco{d,d^*}
$$
is the well-known Laplace-Beltrami operator on $\Om$. 
\vspace{3mm}\par\noindent
{\em Proof of the formula\/} $d=\ep\circ\nabla$. We easily check the validity on 0-forms,
$$
   \ep\circ\nabla f=\ep(dx^i)\pd_if=df,\qquad f\in\Gamma,
$$
then on 1-forms $v=v_idx^i$,
\begin{eqnarray*}
   \ep\circ\nabla v%
&=&\ep(dx^i)\nabla_{\pd_i}v_jdx^j \\
&=&\ep(dx^i)((\pd_i\alpha_j)dx^j-\alpha_j\Gamma^j_{ik}dx^k)\\
&=&\pd_i\alpha_jdx^i\wedge dx^j-\Gamma^j_{ik}\alpha_jdx^i\wedge dx^k\\
&=&d\alpha\qquad\qquad (\mbox{since}\ \Gamma^j_{ik}=\Gamma^j_{ki}),
\end{eqnarray*}
and finally check Leibniz's rule. For $\alpha\in\Om^p$, $\beta\in\Om$,
\begin{eqnarray*}
\ep\circ\nabla(\alpha\wedge\beta)
&=&\ep(dx^i)\nabla_{\pd_i}\alpha\wedge\beta\\
&=&\ep(dx^i)((\nabla_{\pd_i}\alpha)\wedge\beta
    +\alpha\wedge(\nabla_{\pd_i}\beta))\\
&=&(\ep\circ\nabla\alpha)\wedge\beta
    +(-1)^p\alpha\wedge(\ep\circ\nabla\beta)\,.
\end{eqnarray*}
Since this is precisely Leibniz's rule for the exterior derivative $d$,
the proof is complete.
\vspace{3mm}\par\noindent
{\em Proof of the formula\/} $d^*=-\io\circ\nabla$. The check on 0-forms is easiest:
$$
    \io\circ\nabla f =\io(df\otimes 1)=\io(df)1=0\qquad (f\in\Gamma).
$$
Already on 1-forms $v=v_idx^i$, the check is very involved.
The easy part of the calculation is
\begin{eqnarray*}
  \io(\nabla v)&=&\io(dx^i\otimes\nabla_{\pd_i}v_jdx^j)\\
                 &=&\io(dx^i\otimes dx^j)(\pd_iv_j-\Gamma^k_{ij}v_k)
                      \ =\ g^{ij}(\pd_iv_j-\Gamma^k_{ij}v_k).
\end{eqnarray*}
For the harder part we use $d^*=-*d*$ (on $\Om^1$), $*\ep(v)=\io(v)*$, and
introduce the vector field $X$ as the dual of the 1-form $v$: 
$$
        *v=*\ep(v)1=\io(v)*1=\io(v)\om_0=\sigma_X,\qquad X^\#=v\,
$$
Then $*d*v=\DIV X$ since $d\sigma_X=(\DIV X)\om_0$ and $*\om_0=1$.
The remaining task is performed in local coordinates:
\begin{eqnarray*}
\DIV X&=&\pd_j(v_i\dg g^{ij})\dgm \\
      &=&g^{ij}\pd_jv_i+v_i\pd_jg^{ij}+g^{ij}v_i\pd_j\log\dg\\
      &=&g^{ij}\pd_jv_i+v_i(\nabla_{\pd_j}dx^i,dx^j)
         +v_i(dx^i,\nabla_{\pd_j}dx^j)\\
      & &+\frc{1}{4}g^{ik}v_i\pd_k\log\mbox{det}\,g^2\\
      &=&g^{ij}\pd_jv_i-\Gamma^i_{jk}g^{kj}v_i-\Gamma^j_{jk}g^{ik}v_i
         + \frc{1}{4} g^{ij}v_i\pd_j\,\Tr\log\,g^2\\
      &=&g^{ij}(\pd_jv_i-\Gamma^k_{ij}v_k)-
         (\Gamma^j_{kj}-\hf\Tr(g^{-1}\pd_kg)) g^{ik}v_i \\
      &=&g^{ij}(\pd_jv_i-\Gamma^k_{ij}v_k)\,.
\end{eqnarray*}
To obtain the last line we used
$$
   \Gamma^j_{kj}=\hf g^{jl}(\pd_kg_{jl}+\pd_jg_{kl}-\pd_lg_{kj})
                =\hf g^{jl}\pd_kg_{jl}
                =\hf \Tr( g^{-1}\pd_kg)
$$
and as a byproduct we proved:
\begin{equation} 
    dh=\Gamma^j_{ij}dx^i\,,\qquad h:=\log\dg\,.
   \label{log}
\end{equation} 
To find a way to prove the relation $d^*=-\io\circ\nabla$ in the general case,
we return to (\ref{CodDef}) and recast the integrand:
\begin{eqnarray*}
(\alpha,d\beta)&=&(\alpha,\ep(\nabla)\beta)\qquad\qquad(\alpha,\beta\in\Om)\\
&=&(\alpha,\ep(dx^i)\nabla_{\pd_i}\beta)\\
&=&(\alpha,\nabla_{\pd_i}\ep(dx^i)\beta)-(\alpha,[\nabla_{\pd_i},\ep(dx^i)]\beta)\\
&=&-(\nabla_{\pd_i}\alpha,\ep(dx^i)\beta)+\pd_i(\alpha,\ep(dx^i)\beta)
-(\alpha,\ep(\nabla_{\pd_i}dx^i)\beta)\\
&=&-(\io(dx^i)\nabla_{\pd_i}\alpha,\beta)+(\pd_j+\Gamma^i_{ij})(\alpha,\ep(dx^j)\beta)\\
&=&-(\io(\nabla\alpha),\beta)+\io(\nabla X). 
\end{eqnarray*}
Here, we introduced the vector field $X=a^i\pd_i$ with coefficients
$$
                a^i=(\alpha,\ep(dx^i)\beta)\,.
$$
The proof will be complete, once we have shown that the integral
$$  \textstyle
             \int_M \io(\nabla X)\om_0
$$
vanishes (where $\om_0$ is the volume form). The key formula is
$$
              \DIV X =\io(\nabla X)
$$
which is an immediate consequence of (\ref{dvx}) and (\ref{log}).
Note that $X$ is compactly supported if $\alpha$ and $\beta$ are. 
Then (\ref{lx}) allows us to write
$$ \textstyle
   \int_M \io(\nabla X)\om_0=\int_G L_X\om_0 =\int_{\partial G}d\sigma_X =0,
   \qquad  \mbox{supp}(X)\subset G
$$
and to obtain zero with help of (\ref{basn}) and (\ref{stokes}), 
thus completing the proof.

\section{Selfadjoint Dirac Operators}

Armed with these tools we return to our main task which is studying the
relationship of superconnections with Dirac operators. Therefore, let $\E$
be a real Euclidean (or complex Hermitian) superbundle with scalar 
product $(\cdot,\cdot)$ which is also a Clifford module with Clifford action
$c$, and let $D$ be some Dirac operator on $\E$ which means (to remind 
ourselves)
$$
         \sco{D,f}=c(df),\qquad c(df)^2+(df,df)=0\qquad(f\in\Gamma).
$$
We then construct the (formal) adjoint $D^*$ by requiring, for locally
supported $\psi_i\in\G{\E}$,
\begin{equation}
         \int_M(D^*\psi_1,\psi_2)\om_0=\int_M(\psi_1,D\psi_2)\om_0\,.    
  \label{adjD}
\end{equation}
To test whether $D^*$ again is some Dirac operator (for the Clifford action
$c$) we form the supercommutator with (real) functions $f$:
$$
\sco{ D^*,f}=-\sco{ f,D^*}=-\sco{ D,f}^*=-c(df)^*
$$
This clearly shows that $D^*$ fails to be a Dirac operator unless
$$ 
        c(df)^*+c(df)=0\,,
$$ 
that is to say, unless the Clifford module $\E$ is
selfadjoint (see [1] for details concerning selfadjoint modules).
Therefore, the concept of selfadjoint Dirac operators ($D=D^*$) makes sense
only for selfadjoint modules.

In the same manner, if $\nabla^\E$ is a connection on the bundle $\E$, we
would construct the adjoint connection $\nabla^{\E*}$ by requiring
\begin{equation}
  \label{psi}
  d(\psi_1,\psi_2)=(\nabla^{\E*}\psi_1,\psi_2)+(\psi_1,\nabla^\E\psi_2),
  \qquad \psi_i\in\G{\E}.
\end{equation}
This definition automatically makes the Levi-Civita connection selfadjoint:
$\nabla^*=\nabla$.

In particular we find, for a Clifford connection $\nabla^\E$ on a selfadjoint 
module $\E$ and for $\psi_i\in\G{\E}$,
\begin{eqnarray*}
(\psi_1,\nsl^\E\psi_2)&=&(\psi_1,c(dx^i)\nabla^\E_{\pd_i}\psi_2)\\
                      &=&(\psi_1,\nabla^\E_{\pd_i}c(dx^i)\psi_2)
                         -(\psi_1,\sco{\nabla^\E_{\pd_i},c(dx^i)}\psi_2)\\
                      &=&-(\nabla^{\E*}_{\pd_i}\psi_1,c(dx^i)\psi_2)
                         +\pd_i(\psi_1,c(dx^i)\psi_2)\\
                      & &-(\psi_1,c(\nabla^\E_{\pd_i}dx^i)\psi_2)\\
                      &=&(c(dx^i)\nabla^{\E*}_{\pd_i}\psi_1,\psi_2)
                         +(\pd_j+\Gamma^i_{ij})(\psi_1,c(dx^j)\psi_2)\\
                      &=&(q(\nabla^{\E*})\psi_1,\psi_2)+\DIV X
\end{eqnarray*}
where we introduced the vector field $X=(\psi_1,c(dx^j)\psi_2)\pd_j$. Use has
also been made of the quantization map $q=c\circ\sigma^{-1}$.
Integration over $M$ kills the term $\DIV X$. Therefore, and in view of 
(\ref{psi}) we obtain the relation
$$
          q(\nabla^\E)^*=q(\nabla^{\E*})\,,
$$
i.e., selfadjoint Clifford connections $\nabla^\E$ give rise to selfadjoint 
Dirac operators $D=q(\nabla^\E)=\nsl^\E$.

To quote just one example, the Dirac operator $\nsl=d+d^*$ associated to
the Levi-Civita connection is selfadjoint. But to find a similar result for
superconnections, we first have to extend the concept of adjoint actions
in a suitable manner, which implies a passage from $\G{\E}$ to $\Om(\E)$.
Two conditions will do the job:
$$
  df^*=-df,\qquad(\alpha\wedge\beta)^*=\beta^*\wedge\alpha^*,\qquad
  f\in\Gamma,\ \alpha,\beta\in\Om\,.
$$
If $\D$ is a Clifford superconnection and $\nabla^\E$ a Clifford connection,
their difference $L:=\D-\nabla^\E$ is a local operator,
$$     
    L=dx^I\otimes \om_I\in\G{\bw T^*M\skt\End_{C(M)}\E}\,,
$$
where the sum is over subsets $I\subset\{1,\ldots,n\}$.
Granted the above conditions, we have on any selfadjoint Clifford module $\E$
\begin{eqnarray*}
 q(L^*)&=&\om_I^*q(dx^{I*})=\om_I^*q(dx^I)^*\\
       &=&(q(dx^I)\om_I)^*=q(dx^I\otimes\om_I)^*=q(L)^*\,.
\end{eqnarray*}
This way we learn that any Clifford superconnection $\D$ obeys $q(\D)^*=
q(\D^*)$, and, provided $\D=\D^*$ (i.e.\ $L=L^*$), the associated Dirac 
operator $q(\D)$ is selfadjoint. Summarizing:
\begin{quote}\em
There is a 1:1 correspondence between selfadjoint Dirac operators and
selfadjoint Clifford superconnections.
\end{quote}

\section{Complexification}

Let the Riemannian manifold $M$ be even-dimensional and oriented. We consider 
now the space of complex-valued differential forms:
$$
  \Om_\CC=\G{\bw T^*M\otimes\CC}=\sum_{p=0}^n\Om^p_\CC\ ,\qquad
  \Om_\CC^p=\G{\bw^p T^*M\otimes\CC}\,.
$$
As indicated, complex forms arise as sections of the complexified exterior
bundle. Complexification is a necessary step that brings us closer to 
interesting physical applications.

Recall from [1] that on each fiber $\bw T_x^*M\otimes\CC$ we have
an action of the Clifford algebra $C(T^*_xM)$. Therefore, $\Om_\CC$ is a 
Clifford module with Clifford action $c$ extending the Clifford map
$$
               c(v)=\ep(v)-\io(v),\qquad (v\in\V^*).
$$
The extension to $\V^*\otimes\CC$ and $\Gamma(C(M)\otimes\CC)$ is then 
straightforward. Note that complexified spaces inherit complex conjugation
as an involution from the factor $\CC$. Note also that each
fiber $C(T^*_xM)\otimes\CC$ of the complexified Clifford bundle by itself is
a complex Clifford algebra. It is modelled on the complex vector space 
$T^*_xM\otimes\CC$ equipped with the complex-bilinear form $(,)$ extending the 
real-bilinear form $(,)$ in $T^*M$ obtained from the Riemannian structure.

On each fiber $\bw T_x^*M\otimes\CC$ we choose a $\ZZ_2$-grading
given by the chirality of its elements which differs considerably from the
$\ZZ_2$-grading given by $(-1)^p$ where $p$ is the degree of a differential
form. Recall from [1] that the chirality operator is
$$
            \Ga= i^{n/2}e^1e^2\cdots e^n\in C^+(T^*_xM)
$$
where $e^i$ is some oriented (orthonormal) frame of $T^*_xM$, dual to some
frame $e_i$ of $T_xM$. The factor in front guarantees that $\Ga^2=1$.
In terms of sections, the chirality operator may locally be written
$$
  \Ga =i^{n/2}\dg\sigma^{-1}(dx^1\wedge\ldots\wedge dx^n) 
      \in \G{C^+(M)\otimes\CC}
$$
and so is related to the volume form $\om_0$:
$$
    \sigma(\Ga)\equiv c(\Ga)1=i^{n/2}\om_0=i^{n/2}(*1) \in \Om_\CC^n\,.
$$
This formula also indicates a connection between the (linear) chirality 
operator $\Ga$ and the (antilinear) Hodge operator $*$. Indeed, with some 
minor effort one proves
$$
          c(\Ga)\alpha =i^{p(p-1)+n/2}(*\bar{\alpha})\in \Om_\CC^{n-p},
          \qquad \alpha\in\Om_\CC^p
$$
where $\alpha\mapsto\bar{\alpha}$ means complex conjugation.
From $*\ep(v)=\io(\bar{v})*$ it follows that $c(\Ga)\ep(v)=\io(v)c(\Ga)$ and
hence
$$
        c(v)c(\Ga)=-c(\Ga) c(v),\qquad v\in \V^*\otimes\CC
$$
which is the condition necessary for $\Om_\CC$ to be a supermodule with
respect to the parity defining operator $c(\Ga)$.

The Levi-Civita connection $\nabla$ may now be extended so as to remain
a Clifford connection:
$$
      \sco{\nabla,c(a)}= c(\nabla a),\qquad a\in\Gamma(C(M)\otimes\CC)\,.
$$
In particular, we have
$$
  \nabla\sigma(a)=\nabla c(a)1=\sco{\nabla,c(a)}1=c(\nabla a)1=\sigma(\nabla a)
$$
and thus $\nabla\sigma(\Ga)=\sigma(\nabla\Ga)$. The relation may be used to
demonstrate that the chirality operator is ``horizontal'' with respect to the
Levi-Civita connection, i.e., $\nabla\Ga=0$. The result follows from the
representation $\Ga=i^{n/2}\sigma^{-1}(\om_0)$ and the volume form $\om_0$
being horizontal. Locally, we may check the latter assertion by a simple
calculation using (\ref{log}):
\begin{eqnarray*}
\nabla_{\pd_i}\om_0&=&\nabla_{\pd_i}\big(\dg dx^1\wedge\ldots\wedge dx^n\big)\\
&=&\dg\Big((\pd_i\log\dg) dx^1\wedge\ldots\wedge dx^n
                            +\nabla_{\pd_i}(dx^1\wedge\ldots\wedge dx^n)\Big)\\
&=&\Gamma^j_{ij}\om_0
      +\dg\sum_{k=1}^n dx^1\wedge\ldots\wedge\pd_idx^k\wedge\ldots\wedge dx^n\\
&=&\Gamma^j_{ij}\om_0
      +\dg\sum_{k=1}^n dx^1\wedge\ldots\wedge(-\Gamma^k_{ij}dx^j)
          \wedge\ldots\wedge dx^n\\
&=&(\Gamma^j_{ij}-\Gamma^j_{ij})\om_0=0\,.
\end{eqnarray*}
Since $\Ga$ is an even element of $\Gamma(C(M)\otimes\CC)$,
$$
    \sco{\nabla,c(\Ga)}= [\nabla,c(\Ga)]=c(\nabla\Ga)=0\,,
$$
which allows us to rewrite the coderivative:
\begin{eqnarray*}
d^*&=&-\io\circ\nabla=-c(\Ga)\circ\ep\circ c(\Ga)\circ\nabla=
      -c(\Ga)(\ep\circ\nabla) c(\Gamma)\\
   &=&-c(\Gamma)\,d\,c(\Gamma)\,.
\end{eqnarray*}
The standard Dirac operator associated to the Levi-Civita-Clifford connection
may be written in terms of $d$ and $d^*$ as before:
\begin{equation}
  \label{dds}
       \nsl=d+d^*=d-c(\Gamma)\,d\,c(\Gamma)\,.
\end{equation}
For the definition to be correct, we have to check that $\nsl$ is of odd type
(changes the chirality) when applied to $\Om_\CC$. The required property 
follows from the relation $\nsl c(\Ga)=-c(\Ga)\nsl$ which is a consequence 
of (\ref{dds}).

\section{Polarisation}

Many important constructions are based on the cotangent bundle $T^*M$.
Giving the manifold $M$ a Riemannian structure amounts to reducing the
structure group of the cotangent bundle from $GL(n,\RR)$ to $O(n)$. 
Providing an orientation further reduces the group $O(n)$ to $SO(n)$. 
In effect, one is dealing then with the principal $SO(n)$-bundle of oriented
orthonormal frames, $SO(M)$, and its associated vector bundles.
The fiber space of $T^*M$ is $E_n$, the $n$-dimensional oriented Euclidean 
space. Let us assume from now on that $M$ be even-dimensional and $n\ge4$. 
To introduce globally 
the notion of spinors on $M$ we need more: the existence of a principal 
Spin$(n)$-bundle Spin($M$) which doubly covers the bundle $SO(M)$, i.e., 
each fiber of $SO(M)$ is isomorphic to the group 
$\mbox{Spin}(n)=\mbox{Spin}(E_n)$ which doubly covers the group $SO(n)$. 
For instance,
$$
  \mbox{Spin}(4)\cong SU(2)\times SU(2)\,,\qquad \mbox{Spin}(6)\cong SU(4)
$$
Moreover, there should be an isomorphism
\begin{equation}
  \label{spt}
                     T^*M\cong \Spt E_n
\end{equation}
which interprets the cotangent bundle as an associated bundle with fiber
space $E_n$; $M$ is then
said to have a {\em spin structure}. For this to be the case, a necessary
and sufficient condition would be the vanishing of the second Stieffel-Whitney
class $w_2(M)$: true for the sphere $S^n$ but wrong for the complex projective
space $CP(n)$. Similar arguments apply if $M$ carries a semi-Riemannian 
structure. For instance, if the fiber space of $T^*M$ is $M^*_4$, the dual
of the 4-dimensional Minkowski space, the spin group 
$$
          \mbox{Spin}(M^*_4)\cong SL(2,\CC)
$$
doubly covers the Lorentz group $L_+^{\uparrow}$. A Riemannian manifold with
a spin structure is said to be a {\em spin manifold}.

To see the problem of existence from a different perspective we recall 
some results from [1]:
it is always possible to polarize the fiber space $E_n$ of the
cotangent bundle,
$$
  E_n\otimes\CC\cong V\oplus V^*,
  \qquad \mbox{dim\,}V=n/2, 
$$
and then set $S=\bw V$ in order to obtain
$$
  \bw E_n\otimes\CC\cong S\otimes S^*\cong\End S\,, 
  \qquad\mbox{dim\,}S=2^{n/2}.
$$
Recall also that the spinor space $S$ is $\ZZ_2$-graded where parity equals
chirality. Changing the orientation of $E_n$ reverses the chirality.
There may be a topological obstruction to carry out the polarization globally 
and smoothly so as to get a spinor bundle $\mfS$ from $T^*M$. However, given a 
spin structure, the spinor bundle is easily constructed as an associated 
complex vector bundle with an induced $\ZZ_2$-grading:
$$
      \mfS =\Spt S,\qquad \mfS =\mfS^+\oplus\mfS^-\,.  
$$
The dual bundle $\mfS^*$ is introduced in the same manner so as to obtain 
\begin{eqnarray}
  \label{comp}
     \mfS\otimes\mfS^* &=& \Spt(S\otimes S^*)\nonumber\\
     &\cong&\Spt(\bw E_n\otimes\CC)\cong\bw T^*M\otimes\CC
\end{eqnarray}
where the isomorphism (\ref{spt}) was crucial. Since the Clifford algebra
$C(E_n)$ acts on the fiber space $S$ of the spinor bundle, it follows that 
$\mfS$ consists of Clifford supermodules with respect to the Clifford bundle
$$
            C(M) = \Spt C(E_n)
$$
where $g\in\mbox{Spin}(n)$ acts on $a\in C(E_n)$ via the adjoint
representation: $$a\mapsto gag^{-1}\,.$$
In fact, constraints imposed by the condition $w_2(M)=0$ can be relaxed.
We merely need to establish
\begin{itemize}
\item the existence of a spinor bundle $\mfS$,
\item the existence of a complex line bundle $\ell$ such that
  \begin{equation}
    \label{lbu}
      \bw T^*M\otimes_\RR\ell\cong\mfS\otimes\mfS^*\,.\qquad
  \end{equation}
\end{itemize}
The goal can be achieved under less restrictive assumptions. We first explain 
the idea behind the new setting. The Clifford action $c$ on S,
if complexified, leads to an algebraic isomorphism (see [1] for details)
$$
            c:C(E_n)\otimes\CC\to\End S
$$
and, when restricted to $\mbox{Spin}^c(n)\subset C(E_n)\otimes\CC$, reduces to 
a representation
$$
  c:\mbox{Spin}^c(n)\to \mbox{Aut\,}S,\qquad c(g,e^{i\alpha})=e^{i\alpha}c(g)
$$
of the group 
\begin{equation}
  \label{spinc}
  \mbox{Spin}^c(n):=(\mbox{Spin}(n)\times U(1))/\ZZ_2
\end{equation}
where division by $\ZZ_2$ means identifying $(g,e^{i\alpha})$ and 
$(-g,e^{i(\alpha+\pi)})$ in the direct product $\mbox{Spin}(n)\times U(1)$. 
In a sense,
the group $\mbox{Spin}^c(n)$ is the complex analogue of the spin 
group $\mbox{Spin}(n)$. To quote just one example,
$$
    \mbox{Spin}^c(4)\cong \Big(U(2)\times U(2)\Big)\cap SU(4)\,.
$$
The consequence is a commuting diagram with exact rows and columns:
\newcommand{\dar}{\downarrow}
\newcommand{\rar}{\rightarrow}
$$
\def\normalbaslines{\baselineskip30pt\lineskip5pt\lineskiplimit5pt}
\matrix{%
  &       &     &       &1               &          &1       &       &  \cr
  &       &     &       &\dar            &          &\dar    &       &  \cr
 1&\rar&\ZZ_2&\rar&\mbox{Spin}(n)        &\mapr{\mbox{\scriptsize Ad}}   &SO(n)   &\rar   &1 \cr
  &       &     &       &\dar            &          &\dar    &       &  \cr
 1&\rar&\ZZ_2&\rar&\mbox{Spin}^c(n)&\mapr{\pi}&SO(n)\times U(1)&\rar &1\cr
  &       &     &       &\mapd{\delta}   &          &\mapd{} &       &  \cr
  &       &     &       &U(1)            &=\joinrel=&U(1)    &       &  \cr
  &       &     &       &\dar            &          &\dar    &       &  \cr
  &       &     &       &1               &          &1       &       &  \cr} 
$$
where
$$
   \pi(g,e^{i\alpha})=(\mbox{Ad}(g),e^{i2\alpha}),\qquad
   \delta(g,e^{i\alpha})=e^{i2\alpha}\,.
$$
This suggests the following definition:
a $\mbox{spin}^c$ structure on $M$ consists of a principal
$\mbox{Spin}^c(n)$-bundle $\mbox{Spin}^c(M)$ and a principal $U(1)$-bundle 
$P$ together with a double covering
$$
          \mbox{Spin}^c(M)\arr{\pi}SO(M)\times P\,.
$$ 
in a way consistent with the map $\pi$ of the above diagram:
$$
               \pi(pg)=\pi(p)\pi(g),\qquad p\in \mbox{Spin}^c(M),\quad
                                           g\in \mbox{Spin}^c(n)\,.
$$
We submit these data to analysis by constructing the relevant associated 
vector bundles,
\begin{eqnarray*}
               \mfS &=& \Spc S\\
             \mfS^* &=& \Spc S^*\\
               T^*M &=& SO(M)\times_{SO(n)}E_n\\
           \bw T^*M &=& SO(M)\times_{SO(n)}\bw E_n\\ 
               \ell &=& P\times_{U(1)}\CC\,,
\end{eqnarray*}
the point being that we allow the principal $U(1)$-bundle $P$, hence the 
line bundle $\ell$, to be nontrivial. If for some reason it seems possible 
to take $P=M\times U(1)$, we would get the trivial line bundle 
$\ell=M\times\CC$ and so would have effectively introduced a spin structure on $M$. 
Conversely, any spin structure can be extended to a $\mbox{spin}^c$ structure.

A $\mbox{spin}^c$ structure on $M$ gives rise to a group of gauge transformations,
$$
  {\cal G} = \G{\mbox{Ad}\,P},\qquad \mbox{Ad}\,P=P\times_{\rm Ad}U(1)
$$
with $\mbox{Ad}\,P$ the adjoint bundle\footnote{Generally, the adjoint bundle 
is a nontrivial bundle of groups. But here it is isomorphic to $M\times U(1)$ 
because the adjoint action is trivial for an abelian group.}. By construction, 
$\mbox{Ad}\,P$  acts on the line bundle $\ell$. Only in cases where the square 
root $\ell^{1/2}$ exists may we carry this action to the spinor bundle $\mfS$.
Locally, we would then find the following change of the phase of a spinor
field $\psi$:
$$
   \psi(x)\mapsto \pm e^{i\alpha(x)/2}\psi(x),\qquad
   e^{i\alpha}\in\G{\mbox{Ad}\,P}.
$$ 
Consider now the bundle of endomorphisms of $\mfS$ that supercommute with
the Clifford action:
$$
   \mfs=\End_{C(M)}\mfS\,.
$$
Since $C(M)$ acts irreducibly on $\mfS$, the algebra $\mfs^+$, by
Schur's Lemma, consists of scalars only, while $\mfs^-=\{0\}$: 
$$
    \mfs=\mfs^+=\Spc\CC\cong M\times\CC\,.
$$
Since the negative-parity part vanishes, the line bundle $\mfs$, by general 
consent, is said to be ungraded.

\section{Clifford Connections and\\ Superconnections on the Spinor Bundle}

Let us return to the Levi-Civita connection $\nabla$ which has been carried to
various bundles, among them the Clifford bundle $C(M)$. Since $\G{C(M)}$ acts
on $\G{\mfS}$, the space of spinor fields $\psi$, we wish to give a more
explicit description of connections $\nabla^\mfS$ on the spinor bundle $\mfS$ 
satisfying the condition appropriate for a Clifford connection:
$$
   \sco{\nabla^\mfS,c(a)}=c(\nabla a),\qquad a\in\G{C(M)}.
$$ 
Since $\mfS$ is irreducible, two Clifford connections on the spinor bundle
differ by some complex-valued 1-form,
$$
        \nabla^\mfS_1-\nabla^\mfS_2\in\Om^1(\mfs)\,.
$$
To investigate the local structure of $\nabla^\mfS$, we choose a frame field 
$(e_i)_{i=1}^n$ for $TM$, i.e.,
$$
    e_i\in\Gamma_{\rm loc}(TM),\qquad  (e_i,e_j)=\delta_{ij}\,.
$$
In fact, there is a linear relationship between the basis $e_i$ and the basis
$\pd_i$:
$$
    e_i =\br{e_i,dx^k}\pd_k,\qquad \sum_i\br{e_i,dx^j}\br{e_i,dx^k}=g^{jk}\,.
$$
Note that $[e_i,e_k]\ne 0$ in general, while $[\pd_i,\pd_k]=0$ always.
Let $e^i=e^\#_i$ denote the dual basis, i.e.,  
$$
    e^i\in\Gamma_{\rm loc}(T^*M),\qquad  (e^i,e^j)=\delta^{ij}\,.
$$ 
and $\br{e_i,e^k}=\delta_i^k$. Again, there is a linear relationship between
the the basis $e^i$ and the basis $dx^i$:
$$
    e^i=\br{\pd_k,e^i}dx^k,\qquad \sum_i\br{\pd_j,e^i}\br{\pd_k,e^i}=g_{jk}\,.
$$
For the Levi-Civita connection,
$$
    (e_j,\nabla e_k)+(\nabla e_j,e_k)=d\delta_{jk}=0
$$
and, moreover,
$$
    (e_j,\nabla e_k) = \br{\nabla e_k,e^j}=-\br{e_k,\nabla e^j}
   =-(e^k,\nabla e^j)= (e^j,\nabla e^k)\,.
$$
Upon setting $\om_{jk}=(e_j,\nabla e_k)\in\G{T^*M}$ we get 
$\om_{jk}+\om_{kj}=0$ and the following relation in $\G{T^*M\otimes C(M)}$:
$$\textstyle
    \sco{\om_{jk}e^je^k,e^l} = -2\om_{jl}e^j+2\om_{lk}e^k =-4\om_{jl}e^j
    = -4\nabla e^l\,.
$$
Thus $\sco{\om_{jk}e^je^k,a}=-4\nabla a$ for all $a\in\G{C(M)}$.
Consequently, any Clifford connection on $\mfS$ may locally be written
\begin{equation}
  \label{cliff} 
   \nabla^\mfS = d+\hf A-\frc{1}{4}(e_j,\nabla e_k)\,c(e^j)c(e^k)
\end{equation}
where $c$ is the Clifford action on $\G{\mfS}$ and $A$ is some (undetermined)
complex-valued 1-form (or vector potential). We interpret $A$ as the connection 1-form
of a connection $\nabla^\ell$ on the line bundle $\ell$:
$$
               \nabla^\ell=d+A\,.
$$
Reason: if the square root $\ell^{1/2}$ exists,
$\nabla^{\ell^{1/2}}=d+\hf A$. For a $U(1)$ connection on $\ell$, 
the vector potential is purely imaginary:
$$
         A=A_idx^i,\qquad A_i(x)\in i\RR=\mbox{Lie}\,U(1)\,.
$$
This makes $\nabla^\mfS$ a selfadjoint Clifford connection (with respect to
the canonical scalar product in $\mfS$). The structure of the right hand side
of Eq.\ (\ref{cliff}) tells us that the connection $\nabla^\mfS$
respects the grading of $\mfS$:
$$
     \nabla^\mfS : \G{\mfS^\pm}\to\G{\mfS^\pm}\,.
$$      
The associated Dirac operator $D_A=c(e^i)\nabla^\mfS_{e_i}$ on $\G{\mfS}$ may 
be written
\begin{eqnarray*}
    D_A &=&\psl+\hf\asl -\frc{1}{4}(e_j,\nabla_{e_i}e_k)c(e^i)c(e^j)c(e^k)\\
        &=&\psl+\hf\asl-\frc{1}{4}q(\alpha),\qquad
         \alpha=2\alpha_1+3\alpha_3\in\Om^1\oplus\Om^3
\end{eqnarray*}
with $q$ the quantization map. The differential $(1+3)$-form $\alpha$ is 
given by
\begin{eqnarray*}
 \alpha&=&(e_j,\nabla_{e_i}e_k)(e^i\wedge e^j\wedge e^k-\delta^{ij}e^k
        +\delta^{ik}e^j-\delta^{jk}e^i)\\
       &=&(e_j,\nabla_{e_i}e_k)(e^i\wedge e^j\wedge e^k+2\delta^{ik}e^j)\,.
\end{eqnarray*}
Since the Levi-Civita connection is torsion-free, $\nabla_{e_i}e_k-
\nabla_{e_k}e_i=[e_i,e_k]$, and thus
$$
  \alpha_1=\nabla_{e_i}e^i=\br{e_i,\nabla e^i}\in\Om^1,\qquad
  \alpha_3=\frc{1}{3!}(e_j,[e_i,e_k])\,e^i\wedge e^j\wedge e^k\in\Om^3\,.
$$
\vspace{3mm}\par\noindent
{\bf Example}. Assume that $x\in U\subset\RR^n$ are coordinates of local chart
such that $g_{ik}=\lambda^{-2}\delta_{ik}$, where $\lambda(x)$ is some positive
$C^\infty$ function on $U$. This includes two special cases of interest:
$$
\begin{tabular}[t]{lcll}
 (1) the sphere $S^n$ & : & $\lambda=\hf(1+x^2)$ &\qquad $x^2<\infty$\\
 (2) the hyperbolic space $H^n$ & : & $\lambda=\hf(1-x^2)$ &\qquad $x^2<1$
\end{tabular}
$$
The frame field and its dual are given by
$$
            e_i=\lambda\pd_i,\qquad e^i=\lambda^{-1}dx^i\,.
$$
A straightforward calculation yields
$$
     (e_j,\nabla_{e_i}e_k)=(\pd_j\lambda)\delta_{ik}-(\pd_k\lambda)\delta_{ij}
$$
as well as $\alpha_1=(n-1)\lambda^{-1}d\lambda$ and $\alpha_3=0$ so that the 
Dirac operator assumes the following simple form:
\begin{eqnarray*}
    D_A &=& \psl+\hf\asl-\hf(n-1)\lambda^{-1}(\psl\lambda)\\
        &=& \Lambda(\psl+\hf\asl)\Lambda^{-1},\qquad \Lambda^2=\lambda^{n-1}\,.
\end{eqnarray*}
\vspace{3mm}

Let us turn to the square of the Dirac operator for which Lichnerowicz 
provides an explicit expression,
$$
        D_A^2=\Delta^\mfS+\frc{1}{4}r_M+\hf q(F)\,,
$$
with curvature $F$ of the line bundle $\ell$ given by
$$
     F=\hf\sco{\nabla^\ell,\nabla^\ell}\in\Om^2(\ell)\,.
$$       
Locally, $F=\hf F_{ik}(e^i\wedge e^k)$.

$\mbox{Spin}^c$ structures play an important role in ($n=4$)-geometry.
Specifically, the  Seiberg-Witten monopole differential equations [13] 
\begin{eqnarray*}
    D_A\psi &=&0,\qquad\qquad\qquad \psi\in\G{\mfS^+}\\
    F^+   &=&-\frc{1}{4}{\textstyle\sum_{jk}}\br{\psi,c(e^j)c(e^k)\psi}\,
             e^j\wedge e^k
\end{eqnarray*}
have proven to be a powerful tool. In these equations, $F^+=\hf(F+*F)$ denotes
the self-dual part (specific to four dimensions) of the curvature $F$.
For a purely imaginary 2-form $F$, one writes $F\wedge*F=-|F|^2\om_0$, where
$\om_0=*1$ is the volume form, so that locally
$$
            |F|^2=-\hf F^{ik}F_{ik}\ge 0\,.
$$
Likewise, one writes $\br{\psi,\psi}=|\psi|^2$ for $\psi\in\G{\mfS}$ where
$\br{,}$ is the scalar product, taking values in $\G{\mfs}$, induced by the
scalar product $\br{,}$ in $S$ taking values in $\CC$.

The Seiberg-Witten equations imply $|F^+|^2=\frc{1}{8}|\psi|^4$ and minimize 
the action functional
$$
   W(A,\psi)=\int_M\Big(|D\psi|^2+|F^+
   +\frc{1}{4}{\textstyle\sum_{jk}}\br{\psi,c(e^j)c(e^k)\psi}\,e^j\wedge e^k\,|^2\Big)\om_0
$$
which after some manipulations using the formula of Licherowicz may be 
rewritten as
$$
   W(A,\psi)=\int_M\Big(\br{\psi,\Delta^\mfS\psi}+|F^+|^2+\frc{1}{4}r_M|\psi|^2
             +\frc{1}{8}|\psi|^4\Big)\om_0\,.
$$
Note that the Euler-Lagrange equations will come out as second-order 
differential equations for $\psi$ and $A$, while the Seiberg-Witten equations
are of first order.

An elementary modification is achieved by studying Clifford superconnections
$\D$ on the spinor bundle $\mfS$. We will try to determine the extra freedom
obtained this way. From the previous discussion, we already know the general
structure, i.e., $\D =\sum_{p=0}^n\D_p$, where $\D_1$ is a Clifford connection
on $\mfS$ and
$$
      \D_p\in\cases{\Om^p(\mfs^+)=\Om^p(\mfs) & if $p=$odd\cr
                     \Om^p(\mfs^-)=\{0\}       & if $p=$even.\cr}
$$                                           
This shows that there is no room for a Higgs field ($p=0$), and in four
dimensions the extra freedom consists in adding a complex-valued field
which is a differential 3-form.

To be able to incorporate a scalar Higgs field in a Clifford superconnection
we are thus forced to enlarge the spinor module. As candidates we consider 
twisted spinor modules $\mfS\otimes\E$ where $\E$ may be any $\ZZ_2$-graded
complex vector bundle with trivial Clifford action. In a sense, any extension
must be of the twisted type (see [1] for a discussion).

A superconnection $\D$ on $\mfS\otimes\E$ has $p$-forms 
$$
 p\ne1:\qquad \D_p = \one\otimes\Phi_p,\qquad
           \Phi_p\in\cases{\Om^p(\End^-\E) & if $p=$even\cr
                             \Om^p(\End^+\E) & if $p=$odd\cr}
$$
while, for $p=1$, the structure is given by (\ref{cliff}), except that
$c(e^i)$ has to be replaced by $c(e^i)\otimes\one$ and the connection
1-form has to be reinterpreted:
$$
  p=1:\qquad A=\one\otimes\Phi_1,\qquad \Phi_1\in\Om^1_{\rm loc}(\End^+\E)\,.
$$
The construction of the Dirac operator $D=\Dsl$ is then as before. 

The concept of having a $\mbox{spin}^c$ structure may be extended to situations
where one deals with twisted spinor modules. This suggests to replace the 
principal $U(1)$-bundle $P$ by some principal $G$-bundle with structure
group $G\subset U(N)$ (so that $G$ acts on $\CC^N$) and then to define
$$
                  \E =P\times_{G}\CC^N\,.
$$
I seems reasonable to assume that, for fermions in a gauge theory with gauge 
group $G$, twisted spinor modules obtained this way provide the appropriate 
mathematical setting.
\vspace{10mm}\par\noindent
{\Large\bf References}
\begin{enumerate}
\item G.\ Roepstorff and Ch.\ Vehns: An Introduction to Clifford 
Super\-mo\-dules, math-ph/9908029
\item N.\ Berline, E.\ Getzler, and M.\ Vergne: {\em Heat Kernels and Dirac
      Operators}, Springer, Berlin Heidelberg 1992
\item M.F.\ Atiyah and I.M.\ Singer: The index of elliptic operators on
      compact manifolds, Bull.Amer.Math.Soc.\ {\bf 69}, 422 (1963)
\item H.B.\ Lawson, Jr.\ and M.-L.\ Michelson: {\em Spin Geometry}, 
      Princeton Mathematical Series, Princeton Univ.\ Press, Princeton 1989
\item D.\ Quillen: Superconnections and the Chern character, Topology {\bf
      24}, 89 (1985)
\item V.\ Mathai and D.\ Quillen: Superconnections, Thom classes and
      equivariant differential forms, Topology {\bf 25}, 85 (1986)
\item B.\ Thaller: The Dirac Equation, Springer, Berlin Heidelberg 1992
\item J.\ Tolksdorf: The Einstein-Hilbert-Yang-Mills-Higgs action ond the
      Dirac-Yukawa operator, J.Math.Phys.\ {\bf 39}, 2213 (1998)
\item A.\ Connes: Noncommutative Geometry, Academic Press, New York 1994
\item A.\ Connes: The action functional in noncommutative geometry,
      Commun.Math.Phys.\ {\bf 117}, 673 (1988)\\
      A.H.\ Chamseddine and A.\ Connes: A universal action formula,
      hep-th/9606056
\item E.\ Witten: Supersymmetric Yang-Mills Theory on A Four-Manifold,
      J.Math.Phys.\ {\bf 35}, 5101 (1994)
\item J.\ Fr\"ohlich, O.\ Grandjean, and A.\ Recknagel: Supersymmetric
      Quantum Theory and (Non-Commutative) Differential Geometry,
      hep-th/9706132
\item N.\ Seiberg and E.\ Witten: Monopoles, Duality and Chiral Symmetry
      Breaking in N=2 Supersymmetric QCD , Nucl.Phys.\ {\bf 431}, 484 (1994)
\end{enumerate}
\end{document}